\documentclass[12pt]{article}
\usepackage{newtxtext,newtxmath}
\usepackage{dsfont}
\usepackage{graphicx}
\usepackage[letterpaper,margin=1in]{geometry}
\renewenvironment{abstract}
	{\quotation}
	{\endquotation}

\date{}

\makeatletter
\renewcommand{\fnum@figure}{\textbf{Figure \thefigure}}
\renewcommand{\fnum@table}{\textbf{Table \thetable}}
\makeatother

\usepackage{scicite}
\usepackage{url}
\usepackage{xcolor}
\usepackage{bm}
\usepackage{caption}

\def\fig#1{{Fig.~\ref{#1}}}
\def\eq#1{{Eq.~(\ref{#1})}}

\def\scititle{The importance of being discrete: Fluctuations, defects, and density-orientation coupling in agent-based active nematics}
\title{\bfseries \boldmath \scititle} 

\author{
	M.~Dedenon$^{1,2}$,
	C.~Blanch-Mercader$^{3}$,
	K.~Kruse$^{1,2\ast}$,
    J.~Elgeti$^{4\ast}$.\and
	\small$^{1}$Department of Biochemistry, University of Geneva, Geneva \& 1211, Switzerland.\and
	\small$^{2}$Department of Theoretical Physics, University of Geneva, Geneva \& 1211, Switzerland.\and
    \small$^{3}$Institut Curie, PSL Research University, CNRS UMR 168; 75005 Paris, France.\and
    \small$^{4}$
    Theoretical Physics of Living Matter, IBIP and IAS, Forschungszentrum J\"ulich, 52428 J\"ulich, Germany.\and
	\small$^\ast$Corresponding authors. Email: karsten.kruse@unige.ch, j.elgeti@fz-juelich.de\and
}

\begin{document} 
\maketitle

\begin{abstract} \bfseries \boldmath 
We propose an agent-based model of active flexible rods.
Inspired by cytoskeletal flows, we introduce activity by an internal flow that contributes to the dissipative forces.
The active force between our agents is central and reciprocal, ensuring linear and angular momentum conservation.
For nematic activity, we find spontaneous, thresholdless flows and stochastic flow-reorientation, which is accompanied by the formation of topological defects. 
Defects appear and vanish with activity-dependent rates, and $+1/2$-defects self-propel.
These hallmarks of active turbulence are present even on the scale of individual agents. 
The particle-based interactions lead to an emergent coupling between density and orientation that generates density dipoles around $+1/2$-defects.
Finally, we highlight the versatility of our agent-based model by presenting spontaneous flows in three dimensions and tissue growth. 
Our framework opens the way for an integrated description of living materials, including several forms of activity in the same system.
\end{abstract}

\noindent
Living systems are maintained out of thermodynamic equilibrium: their constituents continuously convert chemical energy from the environment into other forms, notably mechanical work. 
For example, suspensions of purified cytoskeletal proteins can exhibit contractile or extensile mechanical stress generated by molecular motor activity~\cite{Backouche2006,Bendix2008,Sanchez2012}.
The constituents of living matter like the cytoskeleton or cell cultures are typically elongated. When they align, these systems can exhibit orientational order on scales larger than the size of the constituents~\cite{Sanchez2012,Gruler2000,Duclos2014}. 
Often, though, there is no preferred front or back, in which case one refers to nematic order. 
As a consequence of orientational order, active stress is generally anisotropic~\cite{Prost2015}. 
Anisotropic active stress plays a central role, for instance, in chromosome segregation~\cite{Brugues2014}, establishment of the anterior-posterior axis in the worm \textit{Caenorhabditis elegans}~\cite{Mayer2010}, or convergence-extension during wing development in \textit{Drosophila melanogaster}~\cite{Etournay2015}. 
Orientational order fields often present singularities, which are called topological defects.
Such defects play an important role in numerous biological processes such as stress organization, shape formation or density accumulation~\cite{Saw2017,Kawaguchi2017,MaroudasSacks2021,Guillamat2022}.

Hydrodynamic theories of living matter describe the dynamics of a small number of coarse-grained physical variables averaging out the discrete nature of constituents~\cite{Prost2015}. 
On the one hand, hydrodynamic equations have a firm basis in symmetries and conservation laws like momentum or density. 
In this phenomenological approach, the link between the coarse-grained material parameters and the microscopic parameters is unknown.
Furthermore, it is unclear on which length scales these continuum descriptions are valid. 
Hallmarks of active nematic fluids such as spontaneous shear flows~\cite{Kruse2004,Voituriez2005,Duclos2018,BlanchMercader2021a} or spontaneous defect unbinding~\cite{Sanchez2012,Dogic2019} were first reported using this type of approach. 

On the other hand, "the importance of being discrete" \cite{Hallatschek2023} can only be uncovered by agent-based models.
In this framework, activity is typically introduced in the form of agent self-propulsion~\cite{Vicsek1995,Joshi2019,Bar2020,Chate2020}, agent turnover~\cite{Basan2011,DeCamp2015,Hallatschek2023}, or explicitly modeled with molecular motors \cite{Nedelec2007,Vliegenthart2020,Yan2022}.
Agent-based models have unveiled long-range orientational order in two-dimensional active systems~\cite{Vicsek1995} and illustrate concepts like homeostatic pressure~\cite{Basan2011}. 
By bridging the gap to smaller length scales, such models allow to test the validity of hydrodynamic descriptions.
Most importantly, they allow one to readily avoid assumptions commonly made in continuum theories, like constant density or homogeneity of activity, and retain the ultimately granular nature of active systems. 
This has led to discoveries like motility-induced phase separation or negative homeostatic pressure~\cite{Tailleur2008,Podewitz2015}.
Yet, few agent-based models conserve momentum, limiting their applicability to specific active systems.
An intermediate framework between continuous and discrete approaches conserving momentum is multi-particle collision dynamics~\cite{Malevanets1999,Shendruk2015}, which was recently applied to active nematic fluids~\cite{Kozhukhov2022}.

In this work we develop a versatile agent-based model for active nematic fluids to uncover the role of granularity, fluctuations, and limits of the hydrodynamic approach.
It features tunable, extensile or contractile active stress, conserves linear and angular momentum, has a proper thermal (i.e. non active) limit, and is compatible with other forms of activity like growth and self propulsion.
Specifically, inspired by cytoskeletal flows, we introduce activity by an internal flow inside the filamentous agents (\fig{fig1}).
Through friction with the neighbors, this generates an extensile or contractile force dipole, depending on the sign of the flow.
Basing our model on earlier agent-based models of tissue growth \cite{Ranft2010,Basan2011}, it accounts for active mechanical stress and can readily be extended to include other, independent forms of activity, such as growth or division. Our framework describes one-component materials or suspensions at high density where fluid-mediated interactions can be absorbed into effective parameters.

We use this model to study the spontaneous emergence of orientational order and flows in active nematic fluids. Remarkably, our simulations show that phenomena uncovered by hydrodynamic analysis can be found down to agent scales.
For example, we observe the emergence of spontaneous flows and self-propulsion of $+1/2$-defects. Furthermore, we explore effects due to the granularity of our system that are not captured by hydrodynamics. Specifically, in channel geometries, we found novel correlations between the nematic orientational field and the flows of agents. The fluctuations naturally present in our simulations lead to dynamic flow fields with bursts of activity and spontaneous creation of topological defects. Also, we report on density variations around defect cores. Finally, we show how this framework can be extended to three dimensions and how to include other active processes present in living systems like growth or self-propulsion.


\subsection*{A multi-particle agent-based model of an active nematic fluid}

We consider $N$ agents consisting each of $P$ particles connected by harmonic bonds into a chain with additional bending rigidity (\fig{fig1}A).
Between agents, short range repulsion accounts for steric effects, while intermediate attraction with cut-off distance $r_c$ mimics, for example, the effects of cross-linking proteins on cytoskeletal filaments or of cell-cell adhesion molecules like cadherins.
In addition, all particles interact via pairwise dissipative and random forces by a dissipative particle dynamics-like thermostat \cite{Warren1997}.
In this framework, linear and angular momentum are conserved (\fig{fig1}B) and the system relaxes to thermal equilibrium in the absence of any other force.
Newton's equation of motion are integrated by an adapted Velocity-Verlet algorithm \cite{Nikunen2003}.
Up to this point, the model is similar to dissipative particle dynamics for solutions of semi-flexible polymers.

We introduce an active force that is inspired by the retrograde actin flow in migrating cells, and ensures nematic symmetry (\fig{fig1}A). Each particle $p$ of an agent $\alpha$ generates an internal active flow with a prescribed velocity $\mathbf{v}_{a,p}$ oriented along the agent axis $\hat{\mathbf u}_{\alpha}=(\mathbf r_{P}-\mathbf r_1)/|\mathbf r_{P}-\mathbf r_1|$ with
\begin{equation}
\mathbf{v}_{a,p}\cdot\hat{\mathbf u}_{\alpha}=v_a\left(2\,\frac{p-1}{P-1}-1\right)
\end{equation}
for $p=1,2,\ldots,P$.
For an activity parameter $v_a>0$, the internal active flow is divergent, whereas it is convergent for $v_a<0$.
This flow is added to the particle's velocity when calculating dissipative forces, resulting in an active force dipole.
The active force applied by particle $q$ on particle $p$ then reads
\begin{equation}\label{eq:fa}
\mathbf{F}^{(a)}_{pq}=-\xi\,\omega(|\mathbf{r}_p-\mathbf{r}_q|)^2\left[\frac{\mathbf{r}_p-\mathbf{r}_q}{|\mathbf{r}_p-\mathbf{r}_q|}\cdot(\mathbf{v}_{a,p}-\mathbf{v}_{a,q})\right]\frac{\mathbf{r}_p-\mathbf{r}_q}{|\mathbf{r}_p-\mathbf{r}_q|},
\end{equation}
where $\xi$ is the inter-agent dissipation coefficient and $\omega(r)$ is a dimensionless weight factor to ensure short-ranged interactions.
As we show below, divergent (convergent) flow generates an active extensile (contractile) stress.
Note also that the active forces are central and have an opposite reaction force such that linear and angular momentum are still conserved (\fig{fig1}B), a key property not retained for example in Vicsek-like models~\cite{Vicsek1995}.
Further details can be found in Supplementary Text with parameters in table~\ref{table:param}.

\subsection*{Spontaneous channel flow}
A hallmark of active nematics is spontaneous flow in a channel~\cite{Voituriez2005,Marenduzzo2007}. 
We demonstrate the power of our model by showing the emergent flow and characterize fluctuations and correlations.

We simulate active agents confined in an infinite channel of width $W$, with perfect slip walls and periodicity $L$ (\fig{fig1}C).
Without activity $v_a=0$, we confirm that the system behaves as a nematic fluid (Supplementary Text, figure~\ref{fig:passive}).
The system is characterized by velocity and nematic tensor fields, $\mathbf{v}(\mathbf{r})$ and $\mathbf{q}(\mathbf{r})$, which are computed by locally averaging over a small area containing $\sim10$ agents the velocities and orientations of individual agents, respectively (supplementary text).
The largest eigenvalue of the nematic tensor field $\mathbf{q}(\mathbf{r})$ is the local nematic order $s_n(\mathbf{r})$, which is zero in the disordered phase and unity in the perfectly ordered phase. 
The corresponding eigenvector is the director field $\hat{\mathbf{n}}(\mathbf{r})$ representing the average orientation of agents in the vicinity of $\mathbf{r}$.
Equivalently, averaging over all agents defines a global nematic order $S_n$ and a global director $\hat{\mathbf{N}}=(\cos\Theta_n,\sin\Theta_n)$ with orientation angle $\Theta_n$, where the nematic symmetry is reflected by limiting $-\pi/2 \leq \Theta_n \leq \pi/2$ (supplementary text).

For extensile activity, $v_a>0$, a shear flow emerges spontaneously (\fig{fig2}A, Movie 1).
The instantaneous flow fields frequently exhibit transient vortices (\fig{fig2}A and figure~\ref{fig:narrow-large}C) and are clearly more complex than simple shear. 
For analysis, we average out these complexities and quantify the linear shear flow by projecting on the first Legendre polynomial with resulting coefficient $V_x$  (supplementary text).
The shear flow fluctuates strongly in time (\fig{fig2}B) and can even occasionally reverse sign.
The stronger the activity and the wider the channel, the more persistent the flow seems to be. 
For narrow channels and low activity, we observe many flow reversals, which become increasingly rare as the channel widens and activity increases
(figure~\ref{fig:vx}).
For the orientation, we find that the local nematic order $s_n$ is largely uniform throughout the channel, and the director field $\hat{\mathbf{n}}$ aligns with the boundaries but fluctuates in space and time (\fig{fig2}C).
Yet, the global director orientation $\Theta_n$ is non-zero and correlates with shear flow direction (\fig{fig2}B,D).

Averaging the shear flow amplitude reveals a transition from a nearly vanishing flow for contractile activity to spontaneous shear flow for extensile activity (\fig{fig3}A). 
The shear flow amplitude grows roughly linearly with activity and with channel width - the latter implies a shear rate that is independent of the channel width.
Notably, shear flows exists even for channel widths for which the system granularity is apparent ($W=15r_c$ corresponding to $\sim2$ agent lengths) (\fig{fig3}A and figure~\ref{fig:narrow-large}A).
While the shear flow dominates, other modes defined by projection on higher order Legendre polynomials in $x$ and Fourier modes in $y$ contribute strongly.
Summing up all other modes shows that activity adds higher order fluctuations to the flow (\fig{fig3}B and figure~\ref{fig:spectrum}).
For this additional fluctuations, contractile and extensile activity seem to have similar effects. 

For extensile activity, our simulations behave as we expect from hydrodynamic theory \cite{Voituriez2005,Duclos2018}.
The agents are well-aligned throughout the system, and tilt of the nematic orientation strongly correlates with the shear flow (\fig{fig3}C,D and figure~\ref{fig:corr}).
On the other hand, contractile activity leads to a break down of global nematic order, and consequentially we observe little average shear flow.
This break down of global order for contractile activity is caused by the nucleation of defects -- see further discussion below and figure~\ref{fig:defects}A.
However, occasionally, we see order and flow appearing in the system. 
In these cases, the flow and orientation are in opposite directions, leading to a negative correlation of the two (figure~\ref{fig:spectrum}D,E).
This is consistent with hydrodynamics where an instantaneous tilt of the director field generates flows, where the direction depends on the sign of the activity.
However in the contractile case, the flows tend to relax the tilt and no persistent flow is obtained.

Overall, the results highlight that our model produces spontaneous flows as described by hydrodynamics.
Yet contrary to the hydrodynamic theory, we do not observe any measurable threshold in activity for the onset of spontaneous flows.
Beyond hydrodynamics, we find transient flows correlated with orientation for contractile activity. 
Furthermore, noise leads to flow reversal at a rate that decreases with the flow amplitude.

\subsection*{Bulk Properties -- Spontaneous flow and bend-instability induced reorientation}
In bulk, the system exhibits a similar spontaneous flow transition as in the channel geometry: for sufficiently strong extensile activity, spontaneous flow emerges (figure~\ref{fig:flow-PBC}).
For $L=30r_c$, we observe extended phases with shear flows that are either oriented in the horizontal or vertical directions (\fig{fig4} and Movie 2). 
Other directions are suppressed by the limited number of available wavevectors in periodic boundary conditions.
These phases persist for a finite time and stochastically switch directions. This is reminiscent of active bursts of reorientation observed in the hydrodynamics of compressible nematics~\cite{Giomi2011}.
During the transition periods, the director field bends and topological defect pairs are created (\fig{fig4}A).

To quantify this behavior, we define a nematic flow tensor that characterizes the alignment of the flow analogously to the nematic orientation tensor (supplementary text).
Here, the degree of nematic order of the flow is measured by $S_v$, while $\Theta_v$ measures the orientation tilt.
The orientation and flow fields are strongly correlated (\fig{fig5}A-C).
"ithout activity, asymmetric viscosity induces a correlation between agent orientation and velocity, as expected.

For extensile activity, the orientation-flow correlation initially decreases with increasing activity, then it increases reaching a maximum around $v_a\sim 3$, and again decreases for even larger activities (\fig{fig5}C).
While the decrease at large activity could be expected from hydrodynamic theory due to active turbulence, the minimum and increase at moderate activity are unexpected.
Closer visual inspection of time resolved orientations (figure~\ref{fig:switches}A-D) indicates that at small activity, the additional activity increases fluctuations, reducing correlations.
At larger activity, the emerging macroscopic flows are strongly correlated to particle orientation.
For contractile activity, the correlation increases reaching a maximum at $v_a\sim -1$, and then decreases for larger activities (\fig{fig5}C).
 
The nematic flow tensor further allows us to characterize the orientational switching of the flow (\fig{fig5}B,D).
The flow exhibits a clear orientation, and than suddenly switches to a different orientation. 
The times between switching events follow an exponential distribution with a characteristic switching time $\tau_{\rm switch}$ (figure~\ref{fig:switches}E-H and supplementary text). 
For small activity ($0<v_a<2$) we observe no switching events during our simulations, indicating stable phases.
For stronger extensile activity, we observe an increase in frequency of switching.

\subsection*{Defect dynamics and density-orientation coupling}
Simulating larger system sizes $L=302r_c$ allow us to study topological defects and flow patterns showing swirls and chaotic behavior characteristic of active turbulence~\cite{Giomi2015,Alert2022} (\fig{fig6}A,B, Movies 3,4 and figure~\ref{fig:N10000}B).
We observe a continuous creation of $\pm 1/2$-defect pairs driven by activity, balanced by annihilation events. 
While for extensile activity ($v_a>0$) we observe the linear increase of density of defects with activity expected from hydrodynamic theory \cite{Thampi2013,Giomi2015}, for contractile activity the density saturates (\fig{fig6}C). 
Note in particular the minimum of defects for zero activity, and the much larger slope for contractile than for extensile activity. 
This asymmetry between contractile and extensile activity was not reported in results obtained by hydrodynamic analysis \cite{Thampi2013,Giomi2015} or multiparticle collision dynamics\cite{Kozhukhov2022}.

As depicted in the inset of \fig{fig6}D, $+1/2$-defects have a polarity $\mathbf{p}=\bm\nabla\cdot\mathbf q/|\bm\nabla\cdot\mathbf q|$.
Together with activity, this asymmetry leads to self-propulsion of $+1/2$-defects~\cite{Sanchez2012}.
In agreement with hydrodynamic theory ~\cite{Pismen2013,Giomi2013,Giomi2014},
defects move opposite to their polarity $\mathbf{p}$ for extensile activity, and along the polar direction for contractile activity (\fig{fig6}D).
Note the asymmetry between extensile and contractile activity with faster self-propulsion in the extensile case for the same amplitude of $v_a$.

One advantage of our particle-based framework is an emerging coupling between density and other fields like the orientation, compared to hydrodynamic theories where one typically has to choose between a multitude of possible couplings~\cite{Kruse2003,Yabunaka2017}. 
In Figure~\ref{fig7}A, we show the relative variation $\Delta\hat{\rho}=[\rho(\mathbf r)-\bar\rho]/\bar\rho$ of the local density $\rho(\mathbf r)$ with respect to the global density $\bar\rho=N/(L_xL_y)$.
We observe strong variations of density, and identify giant number fluctuations for sufficiently large activity (\fig{fig7}B). Again, an asymmetry between extensile and contractile activity emerges.

Close to a defect, we compute the dipole of density variations $d$ at $+1/2$-defect sites, defined in figure~\ref{fig:N10000}E.
For contractile activity, the dipole is typically positive, meaning a region of dilation at the head and a region of compression at the tail of a $+1/2$-defect (\fig{fig7}A,C). 
A histogram reveals that this positive dipole persists even for $v_a=0$ (\fig{fig7}C), while sufficiently extensile activity ($v_a>2$) promotes negative dipole with compression at the head.
More generally, we find that the Hessian $\mathbf H=\bm\nabla\bm\nabla\hat\rho$ of the density is anisotropic (\fig{fig7}D and figure~\ref{fig:N10000}H). Its principal directions align with those of the nematic tensor $\mathbf q$, such that the scalar quantity $\mathbf q:\mathbf H$ is negative on average for $v_a=0$ (Supplementary text, figure~\ref{fig:N10000}I).
In terms of a continuum theory of equilibrium compressible nematics, these findings suggest the existence of a coupling term $f_w=w\,\mathbf Q:\mathbf H$ in the free energy density, where the coupling coefficient $w>0$. This term leads to density dipoles near $+1/2$-defects with the same orientation as in \fig{fig7}C for $v_a=0$ (supplementary text).

We conclude that hydrodynamics describes the propulsion of defects down to the scale of individual agents.
Indeed, we observe very similar trends in defect density and defect-density interactions in much smaller systems and in the channel geometry (figure~\ref{fig:defects}).
However, we find that higher order terms in the free energy of nematic fluids are necessary to capture salient features of our simulations. From the simulations we can furthermore infer the relevant higher order terms.

\subsection*{Scope \& Discussion}
In this work, we present a new theoretical framework to describe active nematic fluids with an agent-based approach. 
We show that incorporation of internal active flows with nematic symmetry at the agent scale gives rise to spontaneous macroscopic flows and self-propulsion of $+1/2$-defects, which are well-known hallmarks of active nematic fluids.
Thus, our framework establishes a correspondence between the mesoscopic scale, where individual agents generate active force dipoles, and the hydrodynamic scale, where activity is captured by a component of the stress tensor.

Activity is not limited to active stress. 
Indeed, cytoskeletal filaments and molecular motors constitute active nematics, while they also grow by assembly. 
Cells in eukaryotic tissues can crawl, generate nematic active stresses, and divide. 
Adding additional features like turnover, active stress, or self-propulsion to hydrodynamic descriptions of active matter is often achieved by introducing new dynamic fields.
These new fields couple to the already existing fields, leading to additional parameters, which can be hard to interpret microscopically.
An example is provided by various propositions for coupling density variations and orientational order~\cite{Adar.2021,Wang.2023,Dedenon.2023}.
Furthermore, symmetry typically allows for different possible expressions for the corresponding coupling terms. 
As a consequence choices have to be made, which are difficult to justify intrinsically. 
Similarly, some agent-based models serve a specific purpose and are sometimes not readily extended to account for additional features. 
In contrast, our framework readily allows for the introduction of active processes beyond active stress as we will show in the following.

Up to this point, we have explored assemblies in two dimensions.
However, an extension to three dimensions is straightforward.
For example, thin layers of active filaments like the actin-cortex of cells are usually treated in two dimensions. 
However, in two dimensions steric interactions are more constraining than in three dimensions. 
To assess the importance of this effect, we simulate the channel geometry presented above with a small but finite thickness $H=6r_c$ (\fig{fig8}A and Movie 5). 
We observe a shear flow pattern similar to \fig{fig2}A, but the flows are less localized to the walls. 
The escape of agents towards the third dimension appears to limit the nematic order and the coherency of the active flows.
This result highlights that even if small compared to lateral dimensions, the third dimension can play an important role.

Furthermore, our model is based on the two-particle growth model~\cite{Basan2011}, which allows us to implement similarly agent growth, splitting and deletion mechanisms with mechanical feedback (supplementary text). 
Inspired by the free growth of cell colonies, we simulate a group of agents growing on a circular patch (\fig{fig8}B and Movie 6). 
Further highlighting the versatility of our framework by readily implementing various boundary conditions, we consider an absorbing boundary such that agents escaping the patch are removed from the simulation.
We observe large domains of uniform nematic order, transiently destabilized through bend deformations and nucleation of $\pm 1/2$-defects (\fig{fig8}B). The outward flow is destabilized such that $+1/2$-defects tend to self-propel as in extensile active nematics (\fig{fig8}B).
Divisions are primarily located at the periphery, as observed for tumor spheroids and in the two-particle growth model~\cite{Montel2012,Podewitz2015}.
These results demonstrate that our model can readily be extended to include further forms of activity. 
Similarly, self-propulsion of filaments~\cite{Duman2018} can be included through tangential driving forces.

Given its versatility, our framework allows us to address a number of open challenges in the field.
1) The difficulty to design well-controlled experimental systems of active matter makes \textit{in silico} experiments provided by agent-based simulations useful to test analytic theories. 
2) Hydrodynamics is valid on length and time scales large compared to individual agents.
Our framework can be used to check if the results from the hydrodynamic theory are preserved on scales relevant to experimental systems.
3) Similarly, our simulations allow us to gain a microscopic understanding of the macroscopic parameters of continuum theory.
4) We highlighted here that a coupling between density and nematic order is relevant. Such couplings between fields can be inferred and quantified using our method. 
5) Discreteness of constituents (cells, filaments, ...) can result in additional effects, even on the macroscopic scale. 
In particular, agent turnover is intrinsically a discrete micro-scale process, which is important to account for~\cite{Hallatschek2023}.
6) Biological systems are inherently noisy. As we have shown above, our framework is capable of capturing noise effects.
7) Biological experiments often display a staggering complexity, ranging from dynamic boundaries, composite systems to multiple sources of activity.
Our framework can handle this complexity, and study the role of different active contributions like agent turnover or self-propulsion.


\begin{figure} 
\centering
\includegraphics[width=0.8\linewidth]{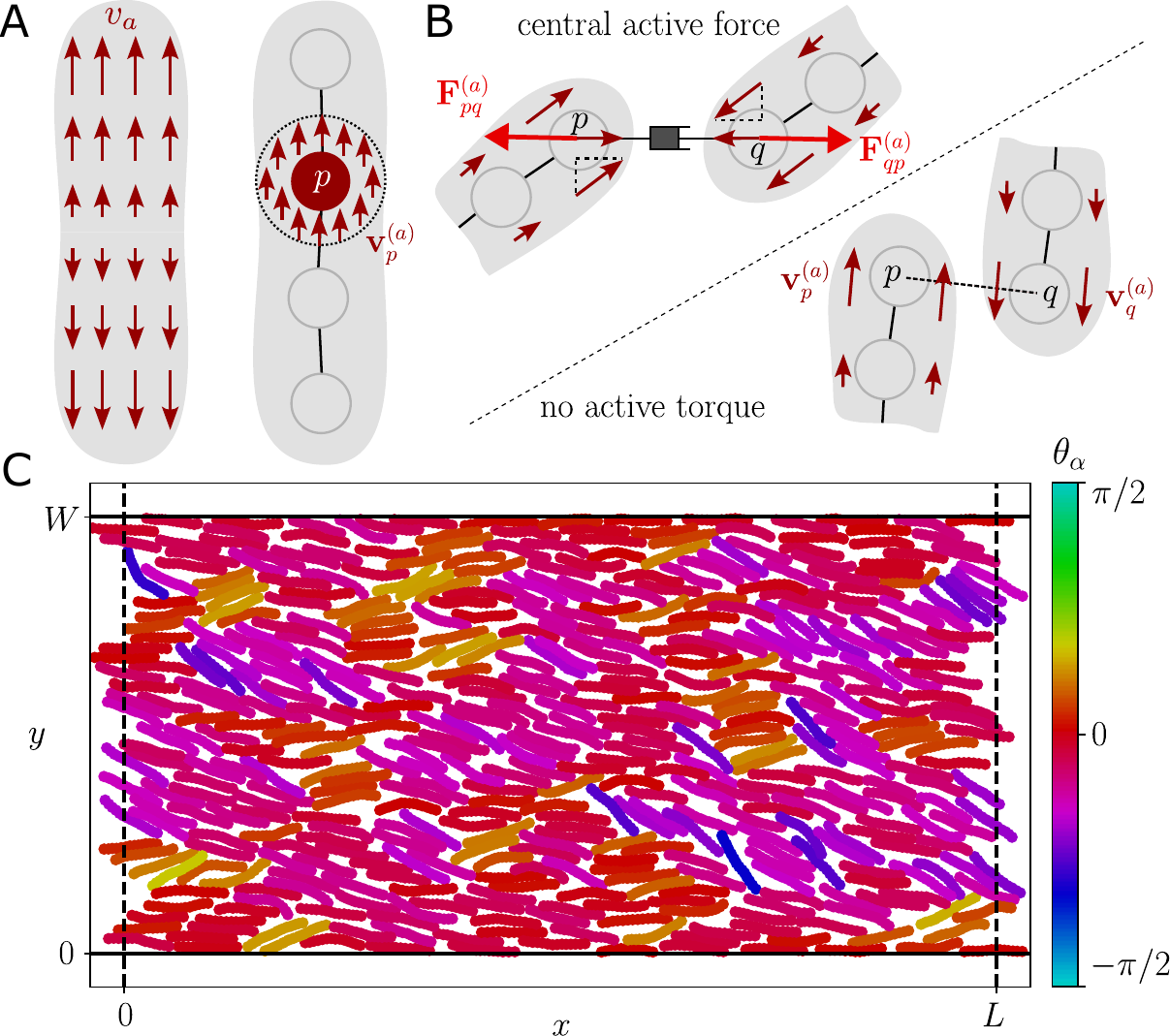}
\caption{\textbf{A multi-particle agent-based model with internal active flows.}
(\textbf{A}) Activity is introduced via an internal flow $\mathbf{v}^{(a)}_p$ along the agent axis, effective over a finite range. Activity is parametrized by the signed amplitude $v_a$, positive for divergent internal flows.
(\textbf{B}) Relative velocities between particles projected on the connecting vector result in a friction force. 
Because the active flow is added to the particle velocities, it results in a reciprocal and central active force.
Consequently, the active force is torque free, 
and vanishes if the connecting vector between particles is orthogonal to the flow.
(\textbf{C}) Snapshot of an active system in the channel geometry at $t=5000$, with periodic boundary conditions at $x=0,L$ and confining wall at $y=0,W$. 
Agents are color-coded according to their nematic orientation, indicated by the angle $\theta_{\alpha}$ with respect to the horizontal axis.
Parameters are $P=14$, $N=450$, $v_a=3$, $t_{\rm sim}=20000$.
}
\label{fig1}
\end{figure}

\begin{figure}
\centering
\includegraphics[width=0.8\linewidth]{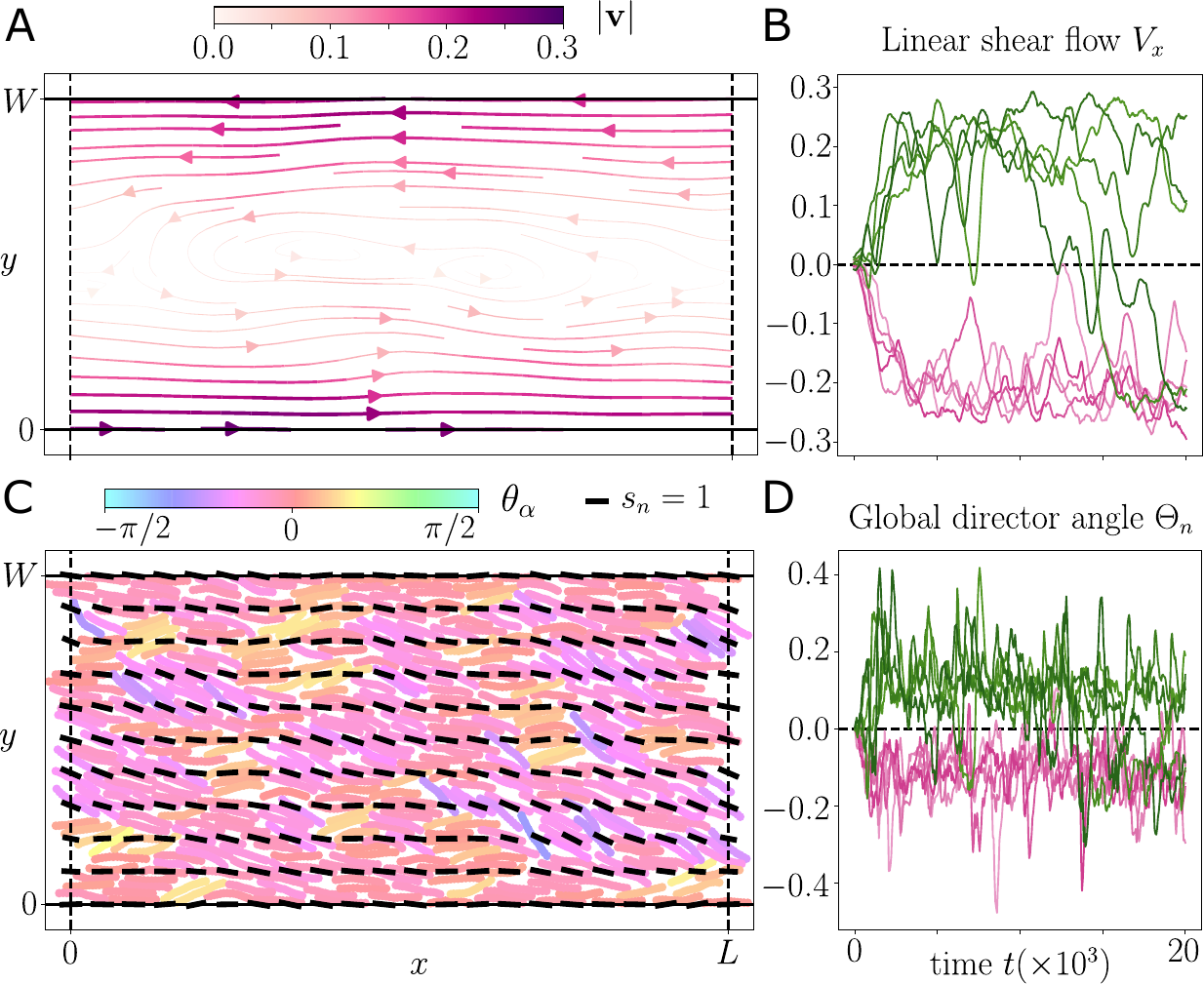}
\caption{Spontaneous channel flow for $v_a=3$.
(\textbf{A}) Coarse-grained velocity field $\mathbf v(\mathbf r)$, at $t=5000$. The colorscale indicates the velocity magnitude $|\mathbf v(\mathbf r)|$. 
(\textbf{B}) Time evolution of linear shear flow component $V_x$ for independent simulations, separating $\langle V_x\rangle_t>0$ (green) and $\langle V_x\rangle_t<0$ (pink) curves.
(\textbf{C}) Coarse-grained director field $\hat{\mathbf n}(\mathbf r)$ (black bars) at $t=5000$, superposed with individual agent positions as in \fig{fig1}C.
The bar lengths indicate the local nematic order $s_n(\mathbf r)$ and the color of individual agents corresponds to their orientation angle $\theta_{\alpha}$.
(\textbf{D}) Time evolution of global director angle $\Theta_n$ for independent simulations. The color code corresponds to (B).
Parameters are $N=450$ with $W=46r_c$ and $L=92r_c$, $t_{\rm sim}=20000$.}
\label{fig2}
\end{figure}

\begin{figure}
\centering
\includegraphics[width=0.9\linewidth]{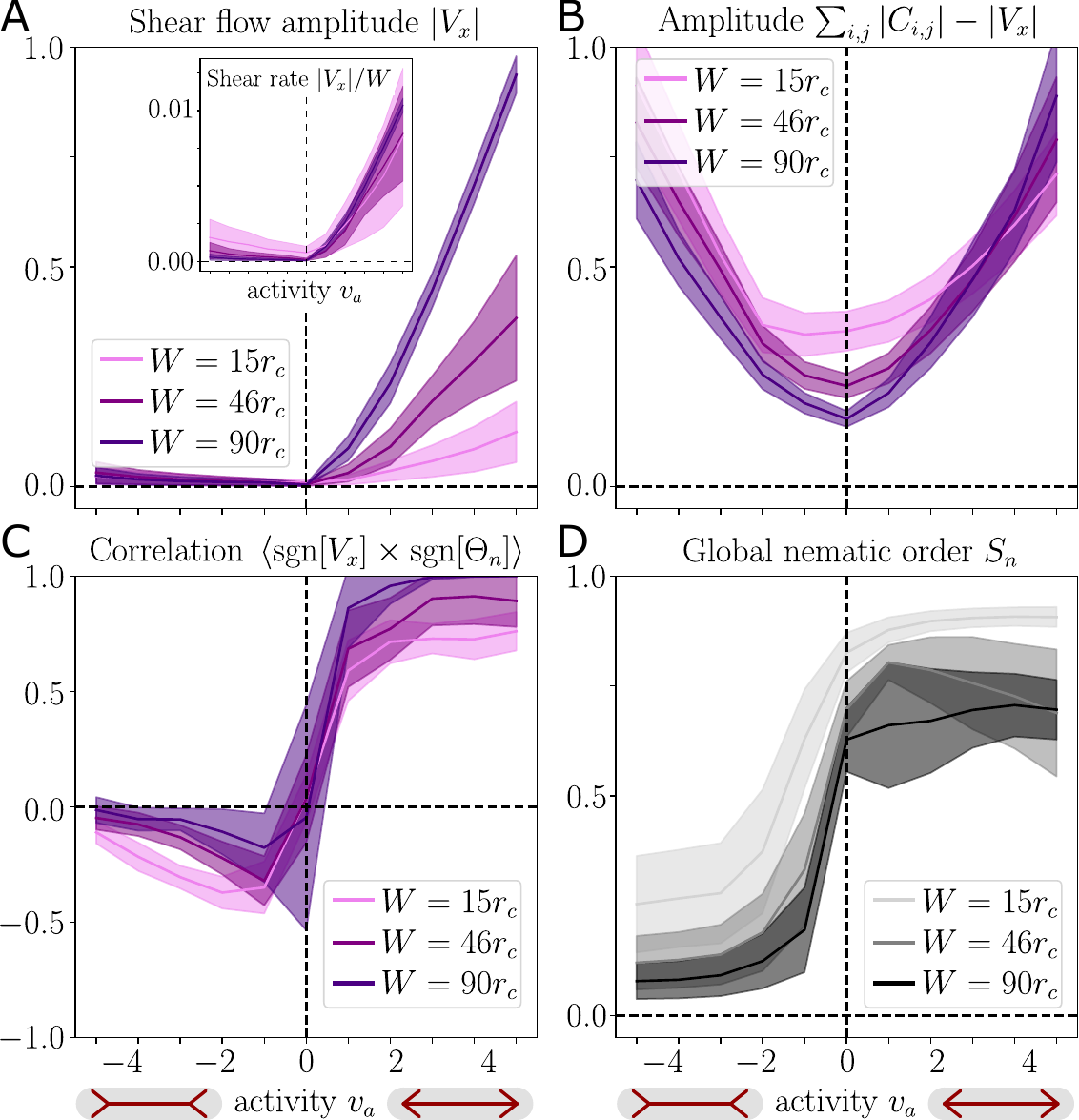}
\caption{\textit{Spontaneous channel flow as a function of activity $v_a$.}
(\textbf{A}) Shear flow amplitude $|V_x|$ as a function of $v_a$, for different $W$. The inset indicates the shear rate $|V_x|/W$.
(\textbf{B}) Non-shear mode contribution of the spectral decomposition as a function of $v_a$, for different $W$.
(\textbf{C}) Sign correlation of shear velocity $V_x$ with global director orientation $\Theta_n$, as a function of $v_a$ and $W$.
(\textbf{D}) Global nematic order $S_n$ as a function of $v_a$ and different $W$, averaged over times and independent realizations.
Parameters are $N=[150,450,900]$ corresponding to $W=[15,46,90]r_c$ and $L=92r_c$, $t_{\rm sim}=20000$, $N_{\rm sim}=20$. Curves indicate mean quantities averaged over independent runs, shaded regions indicate one standard deviation around the mean.}
\label{fig3}
\end{figure}

\begin{figure}
\centering
\includegraphics[width=\linewidth]{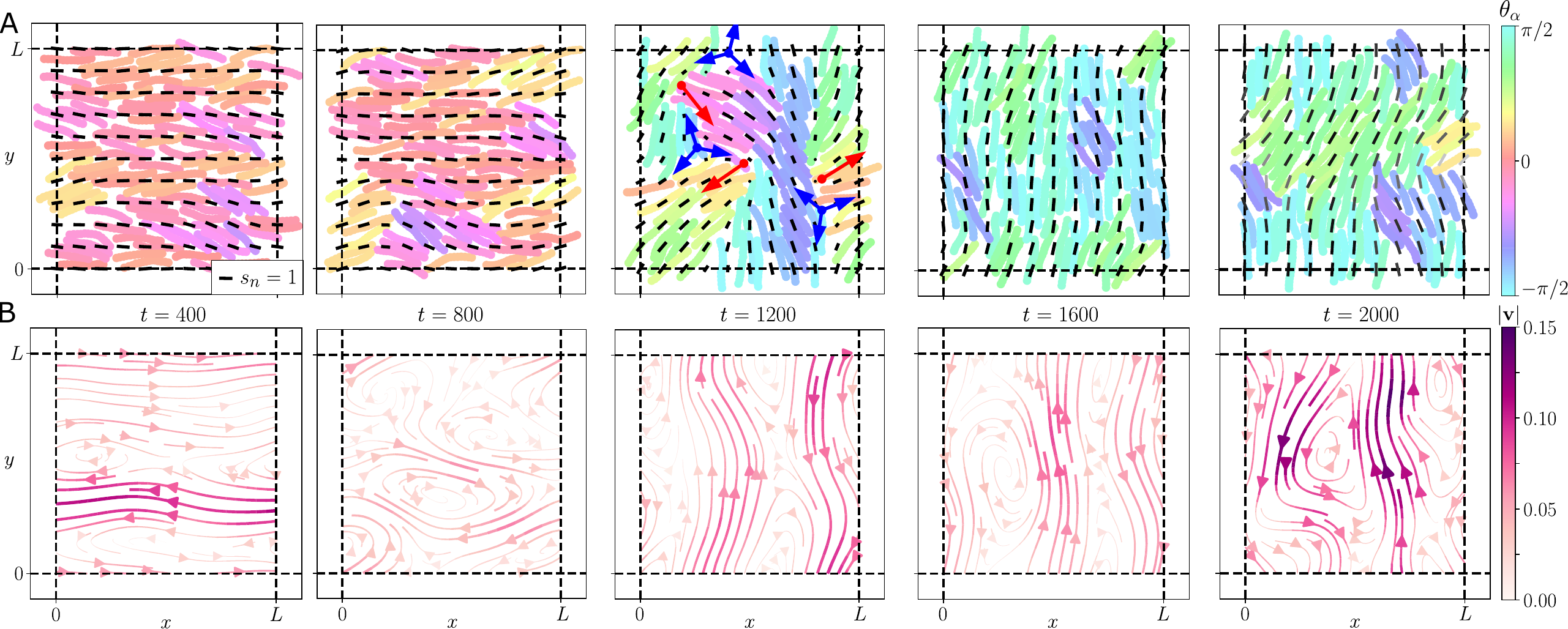}
\caption{\textit{Spontaneous flow in PBCs for $v_a=3$.}
(\textbf{A}) Snapshots of the coarse-grained nematic field superposed with agents positions at different times. Red arrows and blue tripeds indicate the position and orientation of $+1/2$ and $-1/2$ topological defects in the nematic order.
(\textbf{B}) Snapshots of the coarse-grained velocity field for the same times as in A. 
Parameters are $N=100$, $L=30r_c$, $t_{\rm sim}=20000$.
}
\label{fig4}
\end{figure}

\begin{figure}
\centering
\includegraphics[width=0.8\linewidth]{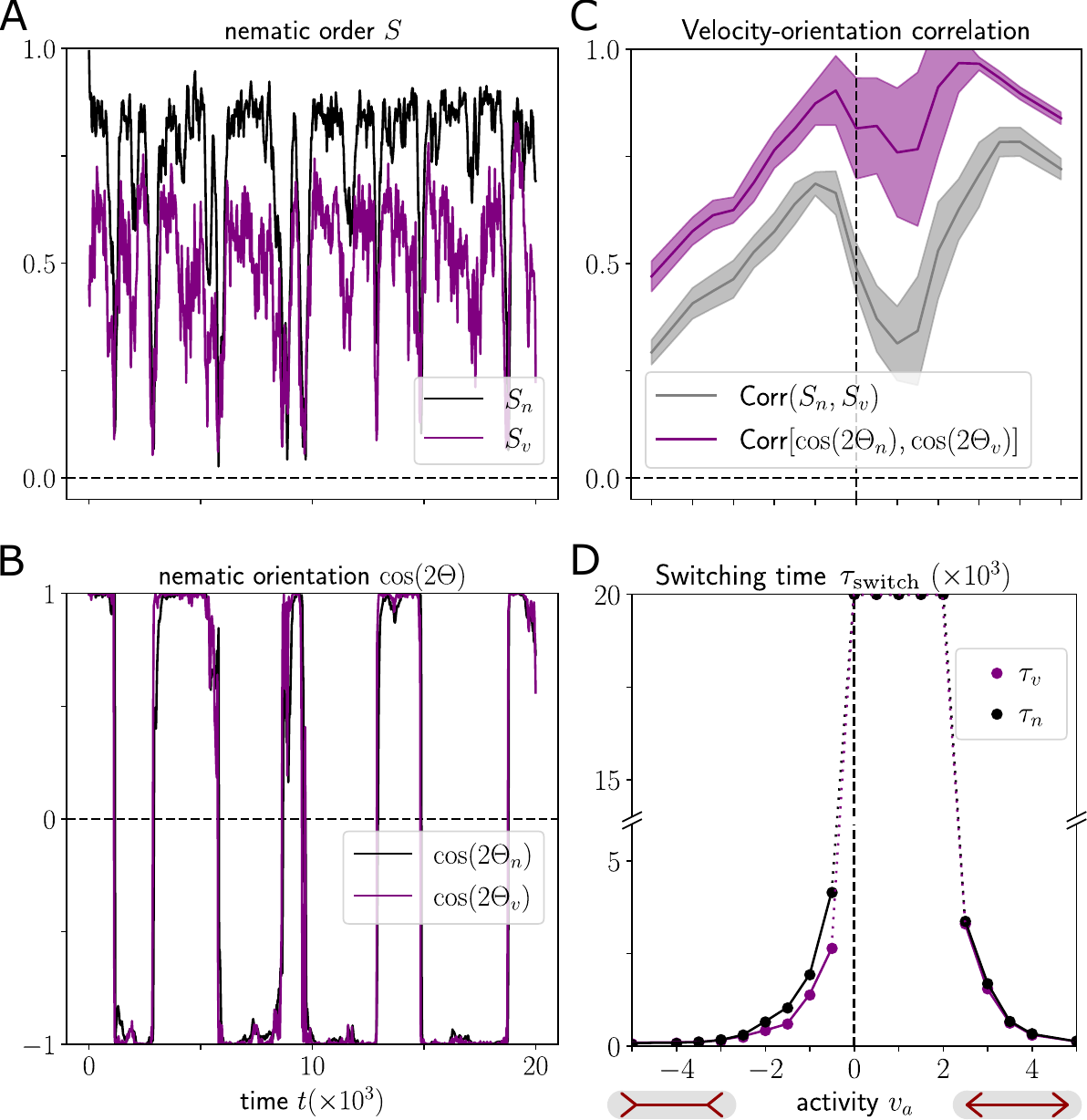}
\caption{\textit{Spontaneous flow in PBCs.}
(\textbf{A,B}) Temporal evolution of nematic order $S$ (A) and global director orientation $\cos(2\Theta)$ (B) obtained from agent orientation in black and agent velocity in purple. Data from the same simulation as in \fig{fig4}.
(\textbf{C}) Instantaneous correlation between $S_n$ and $S_v$ (gray) and between $\cos(2\Theta_n)$ and $\cos(2\Theta_v)$ (purple) as a function of activity $v_a$.
(\textbf{D}) Characteristic switching time $\tau_{\rm switch}$ as a function of activity $v_a$, obtained from a fit of the distributions of switching times from horizontal to vertical velocity alignment (purple), or from horizontal to vertical agent alignment (black). Dashed lines indicate $\tau_{\rm exp}>t_{\rm sim}$. 
In (C,D) parameters are as in \fig{fig4} with $N_{\rm sim}=20$.
}
\label{fig5}
\end{figure}

\begin{figure}
\centering
\includegraphics[width=0.9\linewidth]{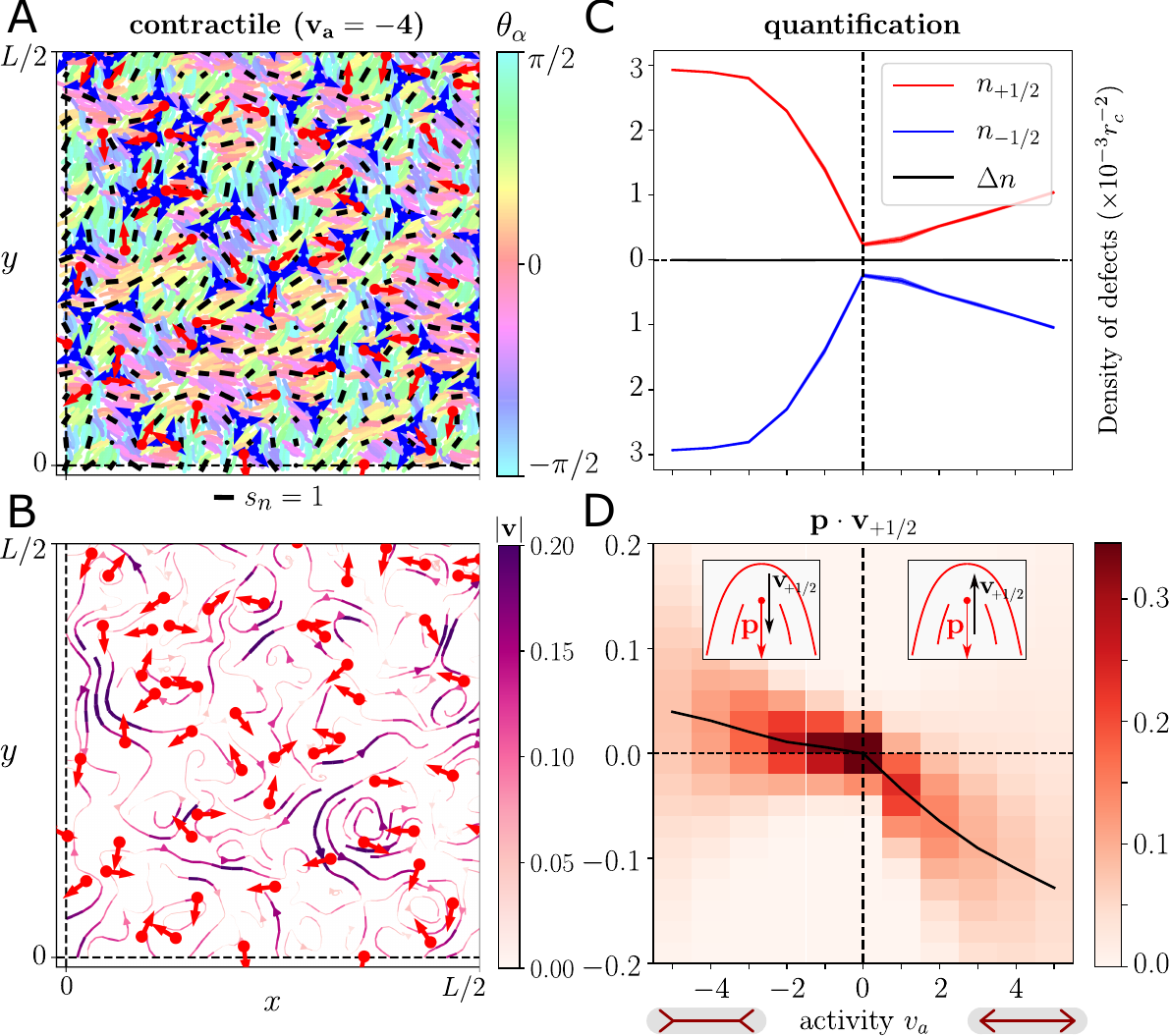}
\caption{\textit{Defect nucleation and self-propulsion.}
(\textbf{A-B}) Snapshots of the periodic system for $v_a=-4$, showing the agent orientations and nematic texture (A), the velocity field (B). 
Red arrows and blue tripeds indicate the position and orientation of $+1/2$ and $-1/2$ topological defects in the nematic order.
(\textbf{C}) Variation of the $+1/2$ (red) and $-1/2$ (blue) defect densities as a function of activity, with net defect charge density $\Delta n=n_{+1/2}-n_{-1/2}$ (black).
(\textbf{D}) Histogram of the velocity $\mathbf v_{+1/2}$ at the $+1/2$-defect core, projected along the defect polarity $\hat{\mathbf p}$ as a function of activity.
Parameters are $N=10000$, $L=302r_c$, $t_{\rm sim}=5000$, $N_{\rm sim}=5$.
}
\label{fig6}
\end{figure}

\begin{figure}
\centering
\includegraphics[width=0.9\linewidth]{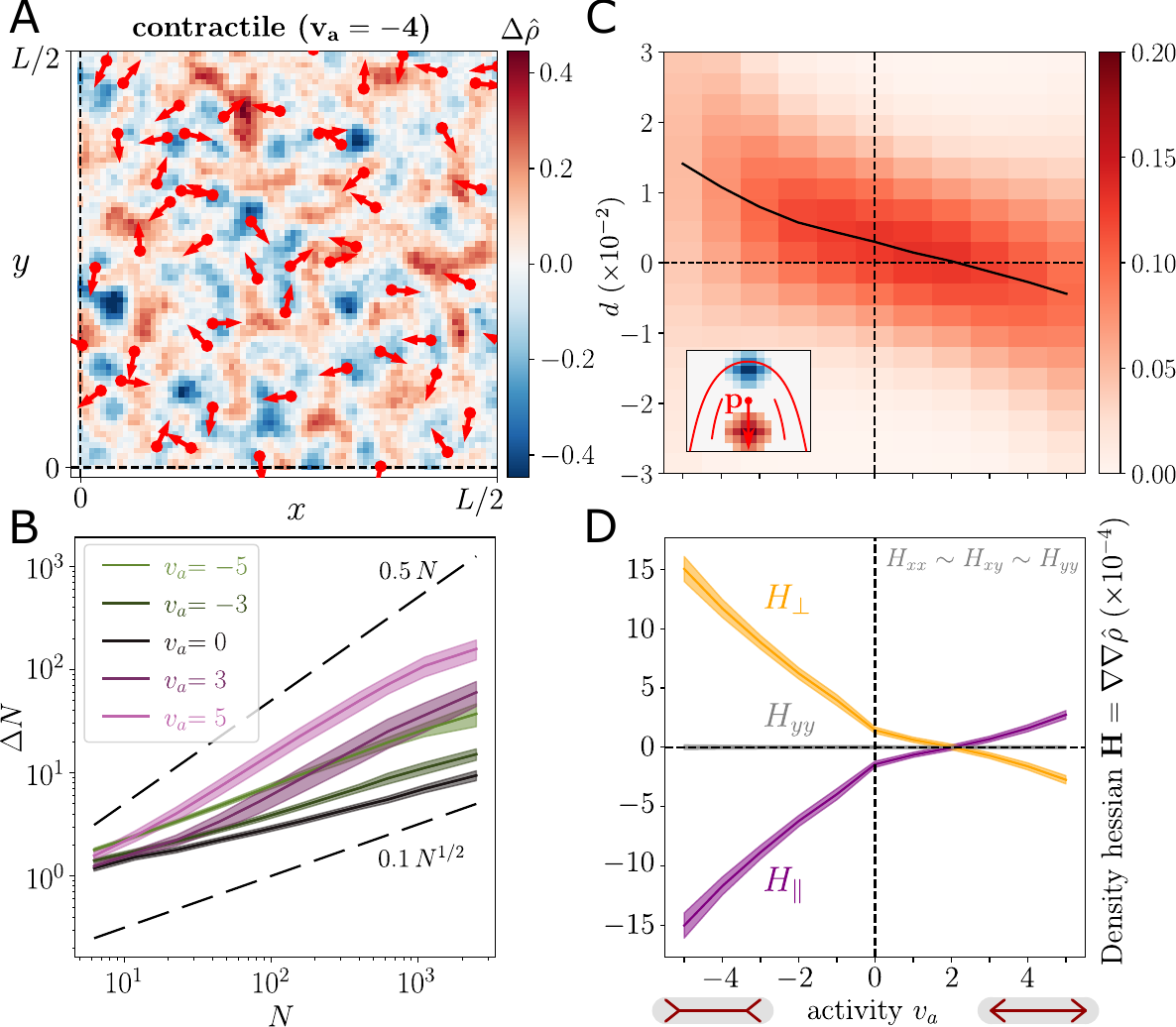}
\caption{\textit{Density-orientation coupling and giant number fluctuations.}
(\textbf{A}) Snapshot of the relative density variations for a periodic system with $v_a=-4$. 
(\textbf{B}) Variance of the number of agents $\Delta N$ for different windows containing $N$ agents on average, for different values of the activity $v_a$.
(\textbf{C}) Histogram of the density dipole $d$ as a function of activity $v_a$, representing the local variation of density along the axis defined by the $+1/2$-polarity $\hat{\mathbf p}$ over a surface patch $A\approx 15\times 15r_c^2$ centered on the defect core position $\mathbf r_{+1/2}$. 
(\textbf{D}) Spatio-temporal average of the components of the Hessian matrix $\mathbf H=\bm\nabla\bm\nabla\hat\rho$, parallel to the nematic director $H_{\parallel}=\langle\hat{\mathbf n}\cdot\mathbf{H}\cdot\hat{\mathbf n}\rangle$ (purple), perpendicular to the nematic director $H_{\perp}=\langle\hat{\mathbf n}_{\perp}\cdot\mathbf{H}\cdot\hat{\mathbf n}_{\perp}\rangle$ (orange), and Cartesian components $H_{ij}=\langle\partial_i\partial_j\hat\rho\rangle$ (gray).
Parameters are $N=10000$, $L=302r_c$, $t_{\rm sim}=5000$, $N_{\rm sim}=5$.
}
\label{fig7}
\end{figure}

\begin{figure}
\centering
\includegraphics[width=.8\linewidth]{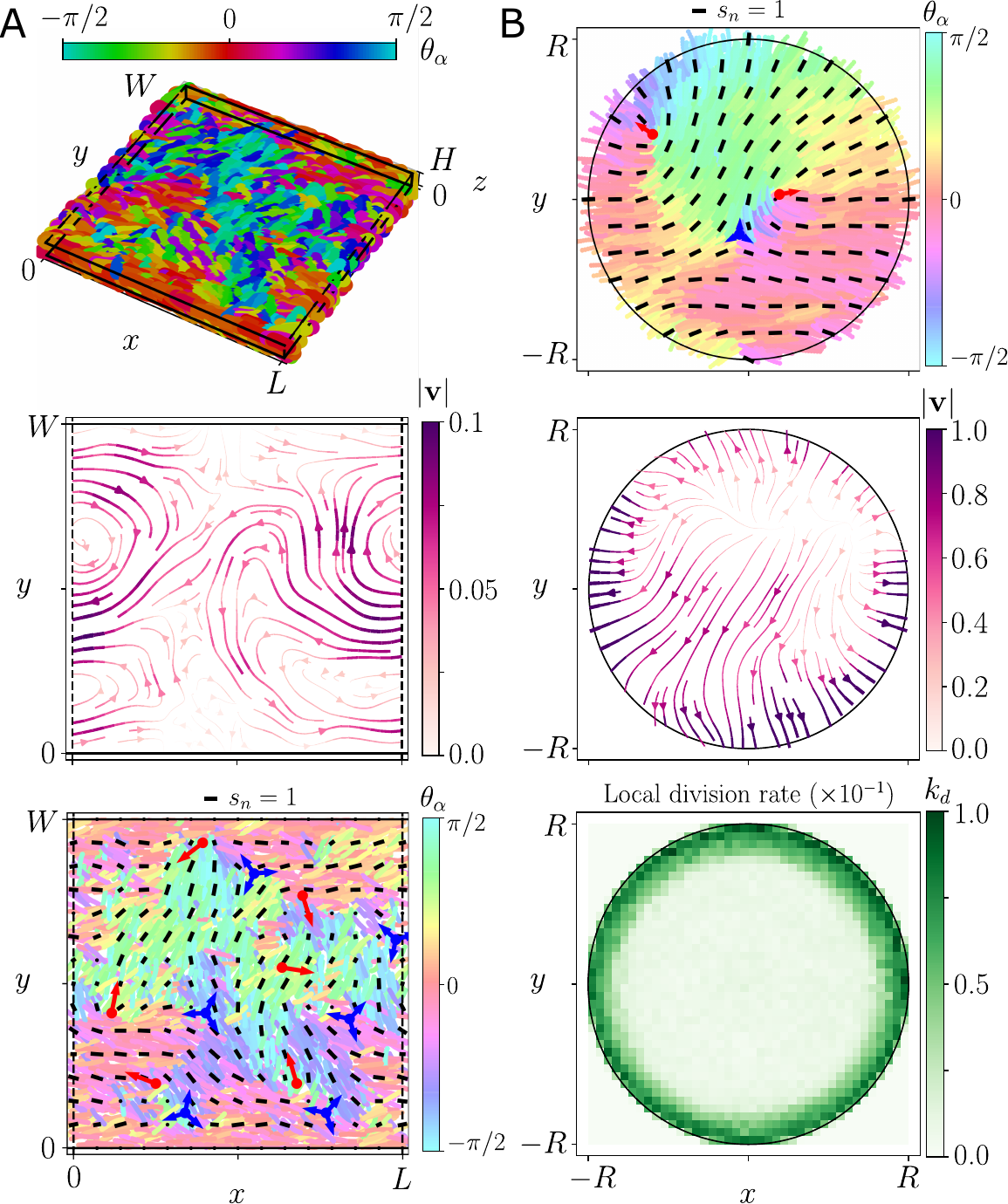}
\caption{\textit{3D flow and 2D tissue growth.}
(\textbf{A}) 3D system under confinement along $y,z$ directions and periodic boundary conditions along $x$, with $L=W=90.5r_c$, $H=6r_c$, $N=3600$, $v_a=2$, $k_{\rm B}T=0.05$.
Top panel indicates a snapshot of particles positions color-coded with respect to $(x,y)$-orientation, middle panel the coarse-grained velocity field projected in the $xy$-plane at $z=H/2$, and bottom panel the coarse-grained two-dimensional nematic field superposed with agents satisfying $|z_{\alpha}-H/2|<H/4$, at $t=1800$.
(\textbf{B}) 2D system with cell divisions and absorbing boundary at radius $R=49r_c$, initiated with $N=1$ agent (current number $N(t)=1198$ at $t=250$) and parameters $v_a=0$, $\xi_s=75$, and $\kappa_b=140$. Top panel indicates a snapshot of particles positions superposed with the nematic texture, middle panel the coarse-grained velocity field, and bottom panel the spatial distribution of the division rate $k_d(\mathbf r)=N_d(\mathbf r)/t_{\rm sim}$. 
}
\label{fig8}
\end{figure}


\clearpage 


\bibliographystyle{sciencemag}

%
%
%
%
%
%


\section*{Acknowledgments}
The computations were performed at University of Geneva on Baobab HPC cluster. We thank Thorsten Auth, Andrew Callan-Jones, Gerhard Gompper, Jean-Fran\c cois Joanny and Daniel Pearce for comments on the manuscript. MD thanks the Forschungszentrum J\"ulich for their hospitality during part of this work.
\paragraph*{Author contributions:}
All authors designed research; M.D. performed simulations and analyzed the data; all authors wrote the paper.
\paragraph*{Competing interests:}
``There are no competing interests to declare.''
\paragraph*{Data and materials availability:}
All data and code needed to evaluate the conclusions in the paper and reproduce the simulations are available at XXX




\subsection*{Supplementary materials}
Materials and Methods\\
Supplementary Text\\
Figs. S1 to S10\\
Table S1\\
References \textit{(36-\arabic{enumiv})}\\ 
Movies S1 to S6\\


\newpage


\renewcommand{\thefigure}{S\arabic{figure}}
\renewcommand{\thetable}{S\arabic{table}}
\renewcommand{\theequation}{S\arabic{equation}}
\renewcommand{\thepage}{S\arabic{page}}
\setcounter{figure}{0}
\setcounter{table}{0}
\setcounter{equation}{0}
\setcounter{page}{1} 


\begin{center}
\section*{Supplementary Materials for\\ \scititle}

M.~Dedenon,
C.~Blanch-Mercader,
K.~Kruse$^{\ast}$\\
J.~Elgeti$^{\ast}$\\
\small$^\ast$Corresponding authors. Email: karsten.kruse@unige.ch, j.elgeti@fz-juelich.de\\
\end{center}

\subsubsection*{This PDF file includes:}
Materials and Methods\\
Supplementary Text\\
Figures S1 to S10\\
Tables S1\\
Captions for Movies S1 to S6\\

\subsubsection*{Other Supplementary Materials for this manuscript:}
Movies S1 to S6\\

\newpage


\subsection*{Materials and Methods}

\subsubsection*{Implementation: algorithmic program structure}
To integrate the equations of motion we use a modified velocity-Verlet algorithm \cite{Nikunen2003}, which accounts for the velocity dependence of forces due to the dissipative interaction.
\begin{align}
\mathbf{v}_p(t+\delta t/2)&=\mathbf{v}_p(t)+\frac{\delta t}{2}\mathbf{F}_p(t) \nonumber \\
\mathbf{r}_p(t+\delta t)&=\mathbf{r}_p(t)+\delta t\,\mathbf{v}_p(t+\delta t/2) \nonumber \\
&=\mathbf{r}_p(t)+\delta t\,\mathbf{v}_p(t)+\frac{\delta t^2}{2}\mathbf{F}_p(t) \nonumber \\
&\rightarrow\text{[division and/or death algorithms]} \nonumber \\
\mathbf{F}^{(\rm c,r)}_p(t+\delta t)&=\mathbf{F}^{(\rm c,r)}_p[\mathbf{r}(t+\delta t)] \nonumber \\
\mathbf{F}^{(\rm d)}_p(t+\delta t)&=\mathbf{F}^{(\rm d)}_p[\mathbf{r}(t+\delta t),\mathbf{v}_p(t+\delta t/2)] \nonumber \\
\bar{\mathbf{v}}_p(t+\delta t)&=\mathbf{v}_p(t+\delta t/2)+\frac{\delta t}{2}\mathbf{F}^{(\rm c,r)}_p(t+\delta t) \nonumber \\
&=\mathbf{v}_p(t)+\frac{\delta t}{2}[\mathbf{F}_p(t)+\mathbf{F}^{(\rm c,r)}_p(t+\delta t)] \nonumber \\
&\rightarrow\text{[iterative loop on $\mathbf v_p$ and $\mathbf F_p^{(\rm d)}$ at $t+\delta t$]} \nonumber \\
&\circlearrowleft\mathbf{v}_p(t+\delta t)=\bar{\mathbf{v}}_p(t+\delta t)+\frac{\delta t}{2}\mathbf{F}^{(\rm d)}_p(t+\delta t) \nonumber \\
&\circlearrowleft\mathbf{F}^{(\rm d)}_p(t+\delta t)=\mathbf{F}^{(\rm d)}_p[\mathbf{r}(t+\delta t),\mathbf{v}(t+\delta t)]
\end{align}
where $\mathbf{F}_p=\mathbf{F}^{(\rm c)}_p+\mathbf{F}^{(\rm d)}_p+\mathbf{F}^{(\rm r)}_p$ and $\mathbf{F}^{(\rm c,r)}_p=\mathbf{F}^{(\rm c)}_p+\mathbf{F}^{(\rm r)}_p$.

To reduce the computational complexity of pairwise forces from $\sim O(P^2N^2)$ to $\sim O(PN)$, we used a standard neighbor list algorithm\cite{Allen2017}. The code is implemented in $C\texttt{++}$ using single thread architecture and simulations were executed on standard CPUs Intel Xeon E5-2630V4 (launch date 2016). Largest simulations with $P\,N=140000$ particles and $t_{\rm sim}=5000$ took $\sim5$ days of computation time, which was reasonable enough to not add other layers of optimization in the code.

\subsubsection*{Implementation: boundary types}
We employ four different boundary types: periodic, free, confining, and absorbing. Periodic and free boundaries are implemented in the usual manner\cite{Allen2017}. For confining boundary, e.g. at $y_w=\pm W/2$, we apply a soft potential $V_w(y)=\frac{1}{2}K_w(y\pm d-y_w)^2$ if $|y|>y_w-d$. The parameter $d=r_c/4$ is a penetration length inside the channel over which the wall potential is non-zero. For absorbing boundary conditions, particles are removed when their center of mass crosses the boundary.

\subsubsection*{Implementation: initial configuration}
The large aspect ratio of agents requires care for the initial preparation of the system when $N_c(t_i)>1$, to avoid agent overlap. Each agent's center-of-mass is assigned a position on a regular grid of length steps $\{a_k\}$ in each direction $k$, such that $a_k=L_k/N_k$. The position is $r_{k,f}=-L_k/2+a_k.\,(f+1/2)$ with $-L_k/2<r_k<L_k/2$ for $f=[0,N_k-1]$. We prepare the system into either isotropic or nematic states.

For an isotropic state, agent orientation $\hat{\mathbf u}_{\alpha}=(\sin\theta_{\alpha}\cos\phi_{\alpha},\sin\theta_{\alpha}\sin\phi_{\alpha},\cos\theta_{\alpha})$ is sampled over uniform random generators ($\phi_{\alpha}=2\pi.\mathrm{rand}[0,1]$, $\theta_{\alpha}=\pi.\mathrm{rand}[0,1]$). One fixes $\theta_{\alpha}=\pi/2$ in two dimensions. Initial agent bond length $l_{0i}$ is reduced to avoid particle collisions, such that the agent length satisfies $l_{0i}.(P-1)<\mathrm{min}(\{a_k\})$.

For a nematic state, its direction is set as $\hat{\mathbf n}=(\sin\theta_n\cos\phi_n,\sin\theta_n\sin\phi_n,\cos\theta_n)$. We only consider nematic directions $\hat{\mathbf e}_x$, $\hat{\mathbf e}_y$ or $\hat{\mathbf e}_z$ for simplicity. Here agents are placed on the defined grid, but shifted alternatively in a chosen orthogonal direction $\hat{\mathbf n}_{\perp}$ by an amount $\pm a_{\perp}/4$. Thus for agent indices $\alpha$ along the nematic direction and $\alpha_{\perp\perp}$ perpendicular to the plane containing $\hat{\mathbf n},\hat{\mathbf n}_{\perp}$ (in three dimensions), one writes $r_{\perp,\alpha}=r_{\perp,\alpha}+a_{\perp}(2\,\alpha\%2-1)(2\,\alpha_{\perp\perp}\%2-1)/4$ where $\%$ is the modulo operation. For elongated agents, the effective transverse spacing is then $a_{\perp}/2$, and one chooses the initial bond length $l_{0i}$ such that $2a_{\parallel}-l_{0i}.(P-1)\sim a_{\parallel}/2$ to have similar agent spacing in the longitudinal direction. This choice has the advantage to reduce the need of initial compaction, and also to start the simulation from a non-crystallized structure. The orientation of each agent $\hat{\mathbf u}_{\alpha}$ follows $\phi_{\alpha}=\phi_n+\Delta\phi.\mathrm{rand}[-1,1]$, $\theta_{\alpha}=\theta_n+\Delta\theta.\mathrm{rand}[-1,1]$ where $\Delta\theta=\Delta\phi=0.1$.

In addition, one assigns for each particle and each spatial component $i$ an initial velocity $v_{p,i}=\sqrt{k_{\rm B}T/m}.\mathcal{N}(0,1)$, where $\mathcal{N}(0,1)$ is a random variable distributed over a normalised centered Gaussian. This means that the initial velocity distribution is Maxwell-Boltzmann and satisfies the equipartition theorem. The net linear momentum per particle is then calculated and subtracted from each $v_{p,i}$ to start with a global system at rest.

\subsubsection*{Implementation: effective aspect ratio and volume}
Using the interaction cut-off range $r_c$, we treat agents as sphero-cylinders (circo-rectangles in two dimensions) of width $r_c$, and length $l_{\alpha}=(P-1)\ell_0+r_c$. 
Thus, the aspect ratio of agents is $\mathrm{ar}=l_{\alpha}/r_c=1+(P-1)\ell_0/r_c$. 
The agent volume is $V_{\alpha}^{(3)}=\frac{4\pi}{3}(r_c/2)^3+\pi(r_c/2)^2\,(P-1)\ell_0$ in three dimensions, and $V_{\alpha}^{(2)}=\pi(r_c/2)^2+r_c\,(P-1)\ell_0$ in two dimensions. 

In two dimensions, we then choose the number of agents $N$ and packing fraction $\mathrm{pf}$ for a simulation, and we obtain the system's size $L=\Omega^{1/2}=(NV_{\alpha}^{(2)}/\mathrm{pf})^{1/2}$. To obtain different sizes in different spatial directions $x,y$, we choose the individual number of agents per dimension such that $N=N_x\,N_y$. Then, we compute a system's length with $L_x=N_x\,(V_{\alpha}^{(2)}/\mathrm{pf})^{1/2}$ whereas the other direction have size $L_y=(N_y/N_x)\,L_x$. This ensures a system's volume $\Omega=L_x\,L_y=NV_{\alpha}^{(2)}/\mathrm{pf}$.

In three dimensions, we start from the two dimensional packing fraction $\mathrm{pf}^{(2)}=N_{\parallel}N_{\perp}V_{\alpha}^{(2)}/(L_{\parallel}L_{\perp})$ in the plane containing the director $\hat{\mathbf n}$, and define the two orthogonal directions $\hat{\mathbf n}_{\perp}$, $\hat{\mathbf n}_{\perp\perp}$ with $N=N_{\parallel}N_{\perp}N_{\perp\perp}$ and $\Omega=L_{\parallel}L_{\perp}L_{\perp\perp}$. To ensure that inter-agent distances are similar in the two directions perpendicular to the nematic direction, one imposes a transverse inter-agent distance $a_{\perp\perp}=a_{\perp}/2$ and define $L_{\perp\perp}=N_{\perp\perp}\,a_{\perp\perp}=(N_{\perp\perp}/N_{\perp})\,L_{\perp}$. Calling $v$ the dimensionless ratio of volumes $v=2V_{\alpha}^{(3)}/\left[V_{\alpha}^{(2)}\right]^{3/2}$, the $3$D packing fraction is written as $\mathrm{pf}^{(3)}=NV_{\alpha}^{(3)}/\Omega=v\,\left[\mathrm{pf}^{(2)}\right]^{3/2}$. For the chosen parameters $\mathrm{pf}^{(2)}$, $r_c$, $P$ and $\ell_0$ (see table \ref{table:param}), one gets $v\simeq 0.57$ and $\mathrm{pf}^{(3)}\simeq 0.41$.

\subsubsection*{Implementation: division and death}
Division of agents is introduced by adding a constant growth force $f_g$ to the shape link force $F^{(\rm l)}(r)=K_l(\ell_0-r)+f_g$. 
This defines an effective link length $\ell_{0g}=\ell_0+f_g/K_l$, and we trigger division when the agent length $l_{\alpha}$ reaches a critical size $\ell_d$ such that $\ell_0.(P-1)<\ell_d<\ell_{0g}.(P-1)$. For an elongated agent of axis $\hat{\mathbf n}_{\alpha}$, division can be performed transversely to the agent axis or longitudinally, and we choose the later case here.

\subsubsection*{Analysis: coarse-grained fields}
From a given configuration of particles $\{\mathbf r_p,\bar{\mathbf v}_p\}$, one builds coarse-grained spatial fields for number density $n(\mathbf r)$, velocity $\bar{\mathbf v}(\mathbf r)$ and nematic order $\mathbf q(\mathbf r)$ over a regular spatial grid of points $\{\mathbf r\}$. This requires the coarse-graining of microscopic distribution functions with a weight function $W(\mathbf R)$
\begin{align}
n(\mathbf{r})&=\sum_p\delta(\mathbf{r}-\mathbf{r}_p)\rightarrow\sum_p\,W(\mathbf{r}-\mathbf r_p) \\ \nonumber
n(\mathbf r)\bar{\mathbf v}(\mathbf{r})&=\sum_p\,\bar{\mathbf v}_p\,\delta(\mathbf{r}-\mathbf{r}_p)\rightarrow\sum_p\,\bar{\mathbf v}_p\,W(\mathbf{r}-\mathbf r_p) \\ \nonumber
n(\mathbf r)\mathbf{q}(\mathbf{r})&=\sum_p\,\mathbf{q}_p\,\delta(\mathbf{r}-\mathbf{r}_p)\rightarrow\sum_p\,\mathbf{q}_p\,W(\mathbf{r}-\mathbf r_p)
\end{align}
The individual nematic tensor $\mathbf{q}_p$ is defined for each particle $p$ belonging to an agent $\alpha$ as $\mathbf q_p=(1/P)(d\,\hat{\mathbf u}_{\alpha}\hat{\mathbf u}_{\alpha}-\mathds 1)/(d-1)$.

The weight function is defined such that $\int\mathrm{d}^d\mathbf R\,W(\mathbf R)=1$, and one chooses for simplicity an isotropic step function $W(R)=\mathrm{If}[R\leq r_w,1/V_d(r_w)],\mathrm{Else}[0]$ parameterized by the window radius $r_w$ and dependent on the $d$-sphere volume $V_d(r_w)$ of the window. The advantage of a step function compared to a smooth (gaussian) kernel is that the spatial integration of the function is not sensitive to the spatial grid resolution. Note that for non-periodic boundaries defined at spatial points $\{\mathbf r_b\}$, $V_d(r_w)$ depends on the distance $|\mathbf r-\mathbf r_b|$ because spatial points external to the system are excluded (precisely to avoid boundary effects). For instance at a boundary point of a flat wall, only half the volume $V_d(r_w)$, internal to the system, must be considered. This effect is captured by a simple linear scaling with the normal distance to boundary $d_n=|\mathbf r-\mathbf r_{b,n}|\leq r_w$ such that $V_d(r_w,\mathbf r)=V_d(r_w).\{\mathrm{If}[d_n\leq r_w,(d_n/r_w+1)/2],\mathrm{Else}[1]\}$.

The window radius $r_w$ is defined from the total number of particles $N_{\rm tot}$ and the system size $\Omega$ such that the window spherical volume $V_d(r_w)$ contains a certain number of particles $N_w$ in the bulk. This imposes the relation $N_w/V_d(r_w)=N_{\rm tot}/\Omega$. One chooses the free parameter $N_w=112$, corresponding to $8$ agents per window or $r_w\simeq 3.8 r_c$ with the particle density considered in results.

\subsubsection*{Analysis: nematic order}
In $d$ spatial dimensions, the global nematic tensor $\mathbf Q$ of a set of $N$ agents with orientations $\hat{\mathbf u}_{\alpha}$ is given by
\begin{equation}
\mathbf Q=\frac{1}{N(d-1)}\sum_{\alpha=1}^{N}\,(d\,\hat{\mathbf u}_{\alpha}\hat{\mathbf u}_{\alpha}-\mathds{1})\simeq\frac{S_n}{d-1}(d\hat{\mathbf N}\hat{\mathbf N}-\mathds{1})
\end{equation}
where $S_n$ is the global nematic order and $\hat{\mathbf N}$ the global director. The second equality assumes a uniaxial nematic, which is expected to be a good approximation for rod-like particles in $3$D, but is exact only in $2$D. Operationally, $S_n=[0;1]$ is the largest positive eigenvalue of $\mathbf Q$ and $\hat{\mathbf N}$ the corresponding eigenvector. Perfect nematic order $S_n=1$ is achieved when $\hat{\mathbf u}_{\alpha}=\hat{\mathbf N}$ for all $\alpha$. From the end-to-end agent displacement $\Delta\mathbf r_{\alpha}=\mathbf r_{\alpha}[P-1]-\mathbf r_{\alpha}[0]$, one defines the agent orientation $\hat{\mathbf u}_{\alpha}=\Delta\mathbf r_{\alpha}/|\Delta\mathbf r_{\alpha}|$. \\
In two dimensions, one can show explicitly that $S_n=\sqrt{(\mathbf{Q}:\mathbf{Q})/2}=\sqrt{Q_{xx}^2+Q_{xy}^2}$ with $\hat{\mathbf N}=\cos\Theta_n\hat{\mathbf e}_x+\sin\Theta_n\hat{\mathbf e}_y$ and $\Theta_n=1/2\,\mathrm{arctan2}(Q_{xy}/Q_{xx})$. Here, the nematic tensor has two degrees of freedom $\{Q_{xx},Q_{xy}\}$, which allows to determine exactly $S_n$ and $\Theta_n$ in the director representation. \\
In three dimensions, the nematic tensor has five degrees of freedom $\{Q_{xx},Q_{xy},Q_{xz},Q_{yy},Q_{yz}\}$ but the director representation only involves three parameters $\{S_n,\Theta_n,\Phi_n\}$ with $\hat{\mathbf N}=\sin\Theta_n[\cos\Phi_n\hat{\mathbf e}_x+\sin\Phi_n\hat{\mathbf e}_y]+\cos\Theta_n\hat{\mathbf e}_z$. This difference originates from the possibility of biaxial nematic order in $3$D, with two principal directions $\hat{\mathbf N}_1$, $\hat{\mathbf N}_2$ with respective orders $S_1$ and $S_2<S_1$ such that $\mathbf{Q}=S_1\hat{\mathbf N}_1\hat{\mathbf N}_1+S_2\hat{\mathbf N}_2\hat{\mathbf N}_2-(S_1+S_2)\hat{\mathbf N}_3\hat{\mathbf N}_3$ where $\hat{\mathbf N}_3=\hat{\mathbf N}_1\times\hat{\mathbf N}_2$. This gives five degrees of freedom with the orthogonality condition $\hat{\mathbf N}_1\cdot\hat{\mathbf N}_2=0$. In that case, one needs to diagonalize $\mathbf Q$, and identify $S_1$ as the largest positive eigenvalue of $\mathbf Q$ with the corresponding eigenvector $\hat{\mathbf N}_1$.

Similarly, one constructs a nematic tensor associated to the particle velocities $\{\mathbf v_p\}$, which reads after nondimensionalization
\begin{align}
\hat{\mathbf Q}_v&=\frac{1}{N_{\rm tot}(d-1)}\sum_{p=1}^{N_{\rm tot}}\,\left(d\,\hat{\mathbf v}_p\hat{\mathbf v}_p-\mathds 1\right),
\end{align}
where $\hat{\mathbf v}_p=\mathbf v_p/|\mathbf v_p|$. In two dimensions, one can write $\hat{\mathbf Q}_v=S_v(2\,\hat{\mathbf N}_v\hat{\mathbf N}_v-\mathds 1)$, with the velocity order parameter $S_v$ and the corresponding velocity director $\hat{\mathbf N}_v=\cos\Theta_v\hat{\mathbf e}_x+\sin\Theta_v\hat{\mathbf e}_y$.

To eliminate the fast velocity fluctuations and focus on large scale coherent motion, one replaces in subsequent analysis the instantaneous velocities $\{\mathbf v_p\}$ by mean velocities $\{\bar{\mathbf v}_p\}$, where $\bar{\mathbf v}_p(t)=\mathbf u_p(t,\overline{\delta t})/\overline{\delta t}$. The temporal displacement $\mathbf u_p$ is forward (backward) for the first (last) time point, and central otherwise with $\mathbf u_p(t,\overline{\delta t})=1/2[\mathbf r_p(t+\overline{\delta t})-\mathbf r_p(t-\overline{\delta t})]$. One chooses $\overline{\delta t}=[1000,10000]\times\delta t=[1,10]$ depending on the data writing frequency.

\subsubsection*{Analysis: spectral decomposition of velocity (periodic boundary conditions)}
To analyse the spatial modes of the velocity field, we perform a Fourier projection over the mean velocities $\{\bar{\mathbf v}_p\}$ at positions $\{\mathbf r_p\}$, for the $N_{\rm tot}$ particles. This is preferred to a discrete Fourier transform on a regular grid, where one would need to compute a coarse-grained velocity field for all acquired times, adding more expensive computations. With a Fourier projection, one can directly use the off-grid information on particles positions.

Each Fourier mode for the velocity component in direction $i$ is defined by a wave-vector $\mathbf k=(k_x,k_y)$, and one obtains complex coefficients 
\begin{equation}
C_i({\mathbf k})=\frac{1}{N_{\rm tot}}\sum_p\,\bar{v}_{p,i}\exp[-\mathrm{i}\,\mathbf r_p\cdot\mathbf k]
\end{equation}
Because particle velocities are real, one has the spectral redundancy $C_i({-\mathbf k})=C_i^*({\mathbf k})$. Note that the zero mode $C_i({\bm 0})$ corresponds to the center-of-mass velocity.
For periodic boundary conditions, the wave-vectors $\mathbf k$ form a discrete set of modes along each spatial dimension $i$, $k_i=2\pi/L_i.\mathbb{Z}$ where $\mathbb{Z}$ is the ensemble of signed integers.

To account for temporal director switches, we also project the velocities along $\hat{\mathbf k}=\mathbf k/|\mathbf k|$ and $\hat{\mathbf{k}}_{\perp}=(-k_y,k_x)/|\mathbf{k}|$, such that the Fourier coefficients are
\begin{align}
C_{\parallel}({\mathbf k})&=\frac{1}{N_{\rm tot}}\sum_p\,(\bar{\mathbf v}_p\cdot\hat{\mathbf k})\exp[-\mathrm{i}\,\mathbf r_p\cdot\mathbf k],\\
C_{\perp}({\mathbf k})&=\frac{1}{N_{\rm tot}}\sum_p\,(\bar{\mathbf v}_p\cdot\hat{\mathbf k}_{\perp})\exp[-\mathrm{i}\,\mathbf r_p\cdot\mathbf k]
\end{align}

An inverse Fourier projection can be defined on a regular grid of spatial points $\{\mathbf r\}$ such that
\begin{equation}
\bar{v}_i(\mathbf r)=\sum_{\mathbf k}\,C_i(\mathbf k)\exp[\mathrm{i}\,\mathbf r\cdot\mathbf k]
\end{equation}
Note that contrarily to a discrete Fourier transform, the projection on the irregular grid of particle's positions $\{\mathbf r_p\}$ implies a loss of information when reconstructing the velocities, hence $\bar{\mathbf v}(\mathbf r)$ can be seen as a parameter-free coarse-grained velocity field with weight function $W(\mathbf R)=1/\Omega\,\sum_{\mathbf k}\cos[\mathbf R\cdot\mathbf k]$ for a system size $\Omega=\Pi_{i=1}^{d}\,L_i$.

For a periodic shear flow $v_x=v_0\sin(2\pi y/L)$ with $k_x=0$, $k_y=k=2\pi n/L$, and $y=[-L/2;L/2]$, using $Y=2\pi y/L$, one gets
\begin{align}
C_{\parallel}(n)&=0, \\ \nonumber
C_{\perp}(n)&=\frac{-1}{N_{\rm tot}}\sum_p\,v_{p,x}\exp[-\mathrm{i}ky_p] \\ \nonumber
&\simeq\frac{\mathrm{i}v_0}{2\pi}\int_{-\pi}^{\pi}\mathrm{d}Y\,\sin(Y)\sin(n\,Y) \\ \nonumber
&=-\mathrm{i}\,\frac{v_0}{\pi}\frac{\sin(n\pi)}{n^2-1}
\end{align}
Thus, one obtains $C_{\perp}(1)=\mathrm{i}\,v_0/2$ and $C_{\perp}(n\neq 1)=0$.

\subsubsection*{Analysis: spectral decomposition of velocity (channel geometry)}
In the channel geometry, due to the confining boundaries, we decompose velocities on an hybrid basis. It is made of Legendre polynomials $P_n(Y)$ in the transverse direction, where $Y=[-1;1]$, and trigonometric functions $\mathrm{exp}[-\mathrm{i}\, xk_x]$ in the longitudinal direction. The complex coefficients are
\begin{equation}
C_i(k_x,n_y)=\frac{2n_y+1}{N_{\rm tot}}\sum_p\,\bar{v}_{p,i}P_{n_y}(2y_p/W)\exp[-\mathrm{i}\,x_pk_x]
\end{equation}
with $n_y\sim\mathbb{N}$ is a positive integer and $k_x=2\pi/L_x.\mathbb{Z}$ as before. An inverse projection can be defined on a regular grid of spatial points $\{\mathbf r\}$ such that
\begin{equation}
\bar{v}_i(\mathbf r)=\sum_{k_x,n_y}\,C_i(k_x,n_y)P_{n_y}(2y/W)\exp[\mathrm{i}\,x k_x]
\end{equation}

For a simple shear flow $v_x=2v_0\,y/W$ with $k_x=0$ and $y=[-W/2;W/2]$, using $Y=2y/W$, one finds coefficients
\begin{align}
C_x(n_y)&=\frac{2n_y+1}{N_{\rm tot}}\sum_p\,v_{p,x}P_{n_y}(2y_p/W) \\ \nonumber
&\simeq v_0(n_y+1/2)\int_{-1}^{1}\mathrm{d}Y\,Y\,P_{n_y}(Y) \\ \nonumber
&=v_0\,\frac{(2n_y+1)\sin(\pi n_y)}{\pi(2-n_y-n_y^2)}
\end{align}
One thus finds $C_x(1)=v_0$ and $C_x(n_y)=0$ for $n_y\neq 1$. A Legendre decomposition is favored here because the Fourier projection of a simple shear flow $v_x=v_0\sin(\pi y/W)$ does not select the pure mode $\mathbf k=(0,\pi/W)$ as additional modes are non-zero, i.e. the Fourier basis is only complete for periodic functions. For instance, one finds $C_x(k_y)=2\mathrm{i}\,v_0\,z\,\cos(z\pi/2)/[\pi(1-z^2)]$ for $k_y=(\pi/W)z$.

\subsubsection*{Analysis: detection of defects in two dimensions}
From the coarse-grained nematic field $\mathbf q(\mathbf r)$ and associated director field $\hat{\mathbf n}(\mathbf r)=(\cos\theta_n(\mathbf r),\sin\theta_n(\mathbf r))$, one computes the winding number field $q(\mathbf r)=(1/2\pi)\oint_{\mathcal{C}(\mathbf r)}\mathrm{d}\theta_n(\mathbf r)$ around a closed loop $\mathcal{C}(\mathbf r)$. All the fields are defined on a discrete grid with $(N_x+1)\times (N_y+1)$ points $\mathbf r_{ij}=(\Delta_x\,i/N_x,\Delta_y\,j/N_y)$ indexed by $i=[0,N_x]$ and $j=[0,N_y]$.
The loop $\mathcal{C}(\mathbf{r})$ is defined as a counter-clockwise nearest-neighbor contour around each grid point $\mathbf r$, such that the displacement map from a point $\mathbf r$ is the set of contour positions $\{\mathbf{r}_c=\mathbf{r}+\mathbf u_c\}$ with $\{u_{c,x}\}=\Delta_x\,\{+,+,0,-,-,-,0,+,+\}$ and $\{u_{c,y}\}=\Delta_y\,\{0,+,+,+,0,-,-,-,0\}$. The contour index $c=[0,N_c]$ identifies first and last positions ($0\equiv N_c$), with $N_c=8$. \\
Then the winding number $q(\mathbf r)$ is computed from an oriented integration of the director angle \cite{Huterer2005}. For each contour position $\mathbf{r}_c$, one defines the angle difference $\Delta\theta_{n,c}=\theta_n(\mathbf r_{c+1})-\theta_n(\mathbf r_c)$ and adds up the total charge $q(\mathbf r)\mathrel{{+}{=}}(1/2\pi)[\Delta\theta_{n,c}+B(\Delta\theta_{n,c})]$ from a loop over $c=[0,N_c]$, where $B(x)=\{\mathrm{If}[x>\pi/2,-\pi],\mathrm{Elif}[x<-\pi/2,\pi],\mathrm{Else}[0]\}$. Finally, one applies an aggregation algorithm to merge topological defects of the same charge which are nearest neighbors on the spatial grid.

For non-periodic boundaries, one cannot define a closed contour to compute a winding number at the wall position. Instead, one defines a half-contour and counts the number of half-rotations, so that $q(\mathbf r)\mathrel{{+}{=}}(1/\pi)[\Delta\theta_{n,c}+B(\Delta\theta_{n,c})]$
with half-contour displacements $\{\mathbf u_{c,+}\}=[\Delta_x,\Delta_y].\{[+,0],[+,+],[0,+],[-,+],[-,0]\}$ and $\{\mathbf u_{c,-}\}=-[\Delta_x,\Delta_y].\{[+,0],[+,+],[0,+],[-,+],[-,0]\}$. The $+$:contour is used for the bottom wall at $y=0$ and the $-$:contour is used for the top wall at $y=W$.

The defect polarity $\hat{\mathbf p}$ of $+1/2$-defects is computed using the unit nematic divergence \cite{Vromans2016} $\hat{\mathbf p}=\bm\nabla\cdot\mathbf q/|\bm\nabla\cdot\mathbf q|$, with derivatives defined from nearest neighbor differences of the field values on the spatial grid. Derivatives are adapted near boundaries (backward/forward instead of central depending on direction) to compute the $+1/2$-defect polarity.

\subsubsection*{Analysis: distribution of switching times}
We consider the evolutions of $|\Theta_n|$ and $|\Theta_v|$ as shown in \fig{fig:switches}A-D. To evaluate the distribution of switching times from horizontal to vertical alignment, we first apply a filter to keep only data points satisfying $(1/4-\epsilon)\pi<|\Theta|<(1/4+\epsilon)\pi$ with $\epsilon=1/10$. This reduces the counting of fluctuation-induced switches. Then, we count the time delay between events such that $|\theta|$ crosses the middle value $\pi/4$. Over the independent simulation realizations, one obtains a set of switching interval $\{\Delta t_s\}$ for each parametric condition.

Then, one computes the cumulant of the distribution $\mathrm{CDF}$ (\fig{fig:switches}E-H) and performs an exponential fit over a time interval $\Delta t_s=[\Delta t_{\rm max}/10:7\Delta t_{\rm max}/10]$, with the maximal switching interval $\Delta t_{\rm max}=\mathrm{MAX}[\Delta t_s]$. This eliminates the discontinuities of the distribution at small and large $\Delta t_s$. We apply the \textit{polyfit} function from the \textit{Numpy} python library at linear order to extract the exponential fitting time $\tau_{\rm exp}$. For parametric conditions where the total number of switching events is below $100$, we consider that the statistics is too weak and assume $\tau_{\rm exp}>t_{\rm sim}$.


\newpage
\subsection*{Supplementary Text}

Here we present:
\begin{itemize}
\item a complete description of the system's equations,
\item the parameters used,
\item the equilibrium limit of our system confined in a channel,
\item the density-orientation coupling in the continuum framework,
\item supplementary figures for Figs.~2,3 in the main text, see Figs.~\ref{fig:narrow-large}-\ref{fig:corr},
\item supplementary figures for Fig.~5 in the main text, see Figs.~\ref{fig:flow-PBC},\ref{fig:switches}.
\item supplementary figures for Fig.~6,7 in the main text, see Figs.~\ref{fig:flow-PBC},\ref{fig:N10000},\ref{fig:defects}.
\end{itemize}

\subsubsection*{Description of the model equations}
\begin{figure}
\centering
\includegraphics[width=0.8\linewidth]{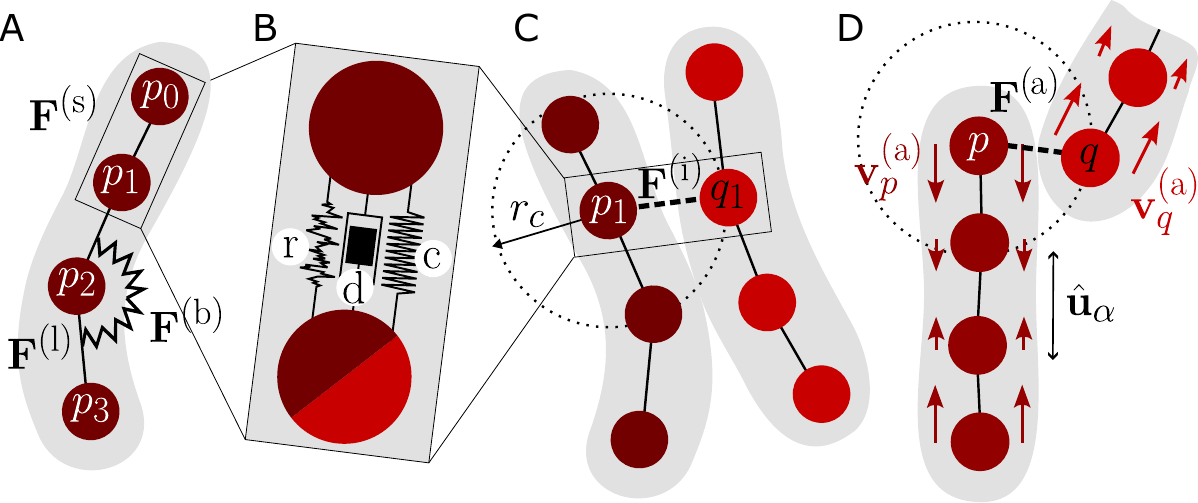}
\caption{\textbf{A multi-particle agent-based model with internal active flows.}
\textbf{A}: Multi-particle agents (here $P=4$ particles) have conservative shape interactions $\mathbf F^{(\rm c,s)}=\mathbf F^{(\rm l)}+\mathbf F^{(\rm b)}$ between intra-agent particles, with link and bend components to ensure agent shape integrity.
\textbf{B}: The interaction between two particles comprises conservative (c), dissipative (d) and random components (r), for both intra-agent and inter-agent cases.
\textbf{C}: Inter-agent forces $\mathbf F^{(\rm i)}$ between particles $p_1$, $q_1$ of different agents are short-ranged with cutting radius $r_c$ (dotted circle).
\textbf{D}: An active force dipole is implemented as an internal treadmilling flow $\mathbf{v}^{(\rm a)}_q$ (convergent here, $v_a<0$) over the particles of each agent, oriented along the agent axis $\hat{\mathbf u}_{\alpha}$ with nematic symmetry. This flow renormalizes the velocities of particles $\mathbf v_p\mapsto \mathbf v_p+\mathbf v^{(\rm a)}_p$ in the dissipative part of the inter-agent force $F^{(\rm d,i)}$, giving an active force contribution $\mathbf{F}^{(\rm a)}$.
}
\label{fig:method}
\end{figure}

We consider $N$ agents that each consist of $P$ particles arranged in a stiff rod, \fig{fig:method}A. Between two particles of the same agent, harmonic links with potential energy 
\begin{equation}
V_{\rm l}\left(\{\mathbf{r}\}\right)=\frac{K_l}{2}\sum_{p=0}^{P-2} (\ell_0-r_{p,p+1})^2
\end{equation}
fix the distance between particles, where $r_{pq}=|\mathbf{r}_{pq}|$ with $\mathbf{r}_{pq}=\mathbf{r}_p-\mathbf{r}_q$ and $\mathbf{r}_p$ is the position of particle p. 
$K_l$ is the spring constant and $\ell_0$ is the equilibrium link length.
A bending energy 
\begin{equation}
V_{\rm b}(\{\mathbf r\})=\frac{\kappa_b}{2\ell_0^3}\sum_{p=1}^{P-2}(\mathbf{r}_{p-1,p}-\mathbf{r}_{p,p+1})^2
\end{equation}
ensures a rod-like shape with bending rigidity $\kappa_b$. The link and bending potential energies contribute to a conservative shape force on particle $p$, $\mathbf F^{(\rm c,s)}_p=-\partial(V_{\rm l}+V_{\rm b})/\partial\mathbf r_p=\mathbf F^{(\rm l)}_p+\mathbf F^{(\rm b)}_p$ as shown on \fig{fig:method}A.

Between two particles from different agents, a conservative force $\mathbf{F}^{(c,\rm i)}_{pq}=F^{\rm i}(r_{pq})\hat{\mathbf{r}}_{pq}$ with $\hat{\mathbf{r}}_{pq}=\mathbf{r}_{pq}/r_{pq}$ accounts for steric repulsion at short distances and attraction at intermediate distances. Specifically, we use
\begin{align}
F^{\rm i}(r)=
 \begin{cases}
      f_0\left((r_c/r)^3-1\right)-f_1 & \text{if $r<r_c$}\\
      0 & \text{otherwise}
    \end{cases} ,
\end{align}
where the constant $f_0$ characterizes the repulsion between two agents, $f_1$ quantifies their attraction, whereas $r_c$ is the cut-off-distance beyond which two particles do not interact, see \fig{fig:method}C. 

For dissipative $\mathbf{F}^{(\rm d)}$ and random $\mathbf{F}^{(\rm r)}$ forces (\fig{fig:method}B), we follow a DPD\cite{Warren1997} mechanism to ensure linear and angular momentum conservation. 
I.e. for two particles  $p$ and $q$,
\begin{align}
\mathbf{F}^{(\rm d)}_{pq}&=-\xi\,\omega(r_{pq})^2[\mathbf{\hat{r}}_{pq}\cdot(\mathbf v_p-\mathbf v_q)]\mathbf{\hat{r}}_{pq}, \label{eq:fd} \\ 
\mathbf{F}^{(\rm r)}_{pq}\mathrm{d}t&=\sqrt{2\xi k_{\rm B}T}\,\omega(r_{pq})\mathrm{d}W_{pq}\mathbf{\hat{r}}_{pq}. \label{eq:fr}
\end{align}
where, $\mathbf{v}_p=\mathrm{d}\mathbf{r}_p/\mathrm{d}t$ is the velocity of particle $p$, $\omega(r)$ is a weight function of distance with $\omega(r<r_c)=1-r/r_c$ or $\omega(r\geq r_c)=0$, $\xi$ has dimensions of a friction constant, $k_BT$ is the effective thermal energy.
The random numbers $\mathrm{d}W_{pq}$ are elementary Wiener processes with zero mean and variance $\langle\mathrm{d}W_{pq}\mathrm{d}W_{pq}\rangle=\mathrm{d}t$. In addition, $\mathrm{d}W_{pq}=\mathrm{d}W_{qp}$ to ensure reciprocity of the interactions. Calling $\delta t$ the simulation time step, elementary Wiener processes are discretized such that $\mathrm{d}W_{pq}=\sqrt{\delta t}\eta_{pq}$, where the random number $\eta_{pq}$ are Gaussian distributed with zero mean and unit variance.
The form of the random and dissipative interaction forces ensure that the (passive) system relaxes to thermal equilibrium~\cite{Warren1997}. 
Furthermore, dissipation and noise can be chosen independently if both particles belong to different agents (inter-agent dissipation coefficient $\xi$) or the same agent (shape dissipation coefficient $\xi_s$ ).
Finally, because all forces are central, that is, along the inter-particle axis, linear and angular momentum are conserved.

The system evolves in time according to Newton's equation of motion (It\^{o}'s convention) for each particle $p$
\begin{align}
m\frac{\mathrm{d}\mathbf{v}_p}{\mathrm{d}t}=&\mathbf{F}^{(\rm ext)}_p+\mathbf{F}^{(\rm c,s)}_p+\sum_{\substack{q\neq p \\ \text{same agent}}}(\mathbf{F}^{(\rm d,s)}_{pq}+\mathbf{F}^{(\rm r,s)}_{pq}) \nonumber \\&+\sum_{\substack{q\neq p \\ \text{different agents}}}(\mathbf{F}^{(\rm c,i)}_{pq}+\mathbf{F}^{(\rm d,i)}_{pq}+\mathbf{F}^{(\rm r,i)}_{pq}+\mathbf{F}^{(\rm a)}_{pq}).\label{eq:forcebalance}
\end{align}
Here we introduced the external force $\mathbf{F}^{(\rm ext)}_p$ which can be used to model, for example, a confining wall or background friction force resulting from interactions with an underlying substrate in the form $\mathbf{F}^{(\rm d,ext)}_p=-\xi_e\mathbf{v}_p$. 
In the latter case, to maintain thermal properties in the passive system, we also add a random force $\mathbf{F}^{(\rm r,ext)}_p\mathrm{d}t=\sqrt{2\xi_e k_{\rm B}T}\,\mathrm{d}\mathbf W_p$. The discretized vector of Wiener processes is $\mathrm{d}\mathbf W_p=\sqrt{\delta t}\bm\eta_p$, where $\bm\eta_p$ is a vector with random components distributed from a Gaussian distribution with zero mean and unit variance.
The active force $\mathbf{F}^{(\rm a)}_{pq}$ is the central part of our work and is detailed below. 
This system of equations is solved by temporal discretization, using a modified velocity-Verlet algorithm~\cite{Nikunen2003} (material and methods). 

Next, we introduce an active force that is inspired by internal cytoskeletal flows \fig{fig:method}D. 
Each particle $p$ of an agent $\alpha$ generates a virtual active flow with a prescribed velocity $\mathbf{v}_{a,p}=v_{a,p}\mathbf{\hat{u}}_{\alpha}$ oriented along the agent axis $\mathbf{\hat{u}}_{\alpha}$. 
The agent axis is defined as the normalized end-to-end vector $\Delta\mathbf r_{\alpha}=\mathbf r_{\alpha}[P-1]-\mathbf r_{\alpha}[0]$, and $\hat{\mathbf u}_{\alpha}=\Delta\mathbf r_{\alpha}/|\Delta\mathbf r_{\alpha}|$.

The active flow profile is
\begin{equation}
v_{a,p}=v_a\,\left(\frac{2p}{P-1}-1\right)
\end{equation}
for $p=[0,P-1]$, with an amplitude equal to $v_a$ at the outer-most particles, and decreasing linearly to zero towards the center. 
For $v_a>0$, this flow points outwards, and for $v_a<0$, the agent generates internal convergent flows.
These flows result in an active force, by including them in the inter-agent dissipation as in Eq~\ref{eq:fd}. 
That is, the force of particle $q$ on particle $p$ in different agents generated by this active process reads 
\begin{equation}
\mathbf{F}^{(a)}_{pq}=-\xi\,\omega(r_{pq})^2[\mathbf{\hat{r}}_{pq}\cdot(\mathbf{v}_{a,p}-\mathbf{v}_{a,q})]\mathbf{\hat{r}}_{pq},
\end{equation}
which has the same form as the dissipative interaction forces $\mathbf{F}^{(d)}$ in \eq{eq:fd}. 
The parameter $\xi$ is the dissipative coefficient from inter-agent interactions.

\subsubsection*{Parameters}
The parameters are summarized in Table~\ref{table:param}, 
where the units are chosen such that $m=r_c=1$ and $\delta t=10^{-3}$. 
To limit the influence of inertia on the system, parameter combinations have been chosen such that inertial time scales are smaller than other relaxation time scales. 
Link stiffness $K_l$ and bending rigidity $\kappa_b$ are chosen large enough to obtain an almost inextensible rod-like shape.

In this work, we choose a parameter set for which the system is in a nematic fluid phase at equilibrium, $v_a=0$. Specifically, we choose a number of particles per agent $P=14$ (i.e. particles of aspect ratio 7), a packing fraction $\mathrm{pf}=0.8$ (material and methods), and a temperature $k_{\rm B}T=0.1$. The mean-squared displacement and the nematic orientation of the system varies with the packing fractions and with the temperatures as expected for other models of nematic liquid crystals~\cite{Onsager1949,Maier1958}, see \fig{fig:passive}. Furthermore, unless otherwise stated, the initial condition corresponds to a set of evenly distributed agents that are aligned in the same direction, which is typically horizontal (materials and methods).

\subsubsection*{Equilibrium limit for a channel geometry}

\begin{figure}
\centering
\includegraphics[width=.95\linewidth]{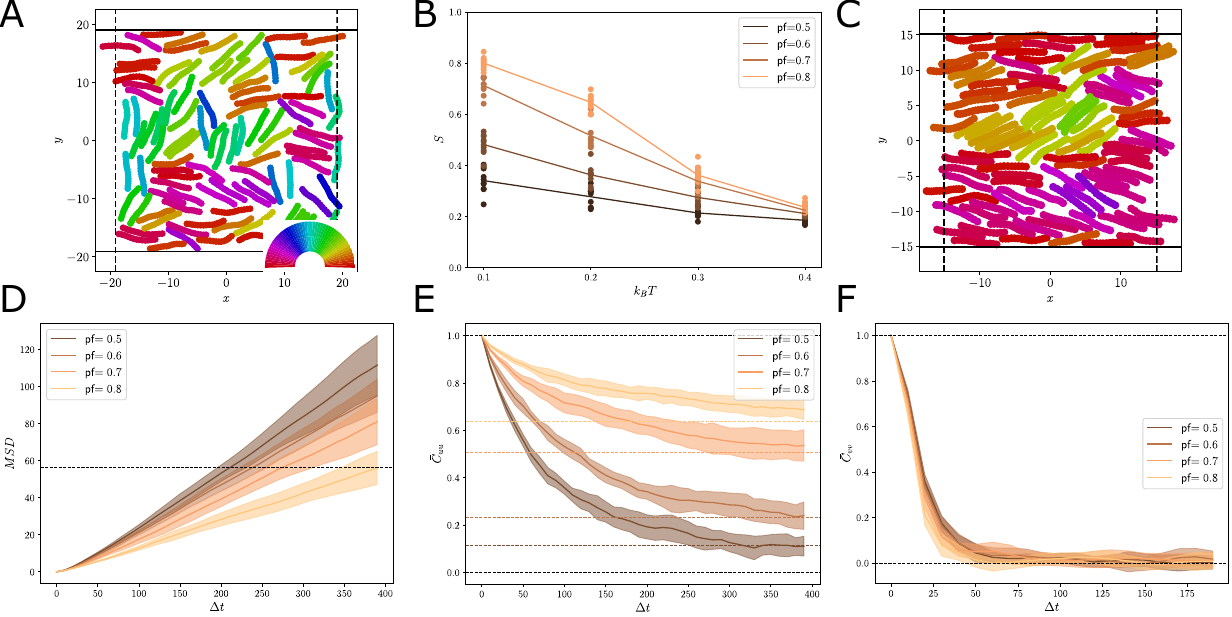}
\caption{\textit{Supplementary results for a passive system in channel geometry, with $P=14$, $N=100$, $t_{\rm sim}=2000$.}
\textbf{A,C}: Snapshot of the system at last simulation time point $t=2000$, for $\mathrm{pf}=0.5$ (A) or $\mathrm{pf}=0.8$ (C) with $k_{\rm B}T=0.1$.
Agents are colored depending on their orientation, see the inset in (A) for the nematic color wheel.
\textbf{B}: Global nematic order $S$ as a function of temperature $k_{\rm B}T$ and packing fraction $\mathrm{pf}$. Dots represent time-averaged $S$ from each simulation. 
\textbf{D-F}: Agent mean-square displacement $MSD$ (D), agent orientation auto-correlation function $\bar{C}_{uu}$ (E) and agent velocity auto-correlation function $\bar{C}_{vv}$ (F) as a function of time difference $\Delta t$, varying $\mathrm{pf}$ at fixed $k_{\rm B}T=0.1$. The dashed line on (D) indicates the square of the agent length $l_a=r_0(P-1)+r_c$. The color dashed lines on (E) indicate average $S^2$ values for each condition. \\
Averages are performed over $N_{\rm sim}=10$ independent simulations. Averages are indicated by full lines (B,D,E,F), and filled regions represent deviations from the mean of one standard error (D-F).
}
\label{fig:passive}
\end{figure}

\begin{figure}
\centering
\includegraphics[width=.85\linewidth]{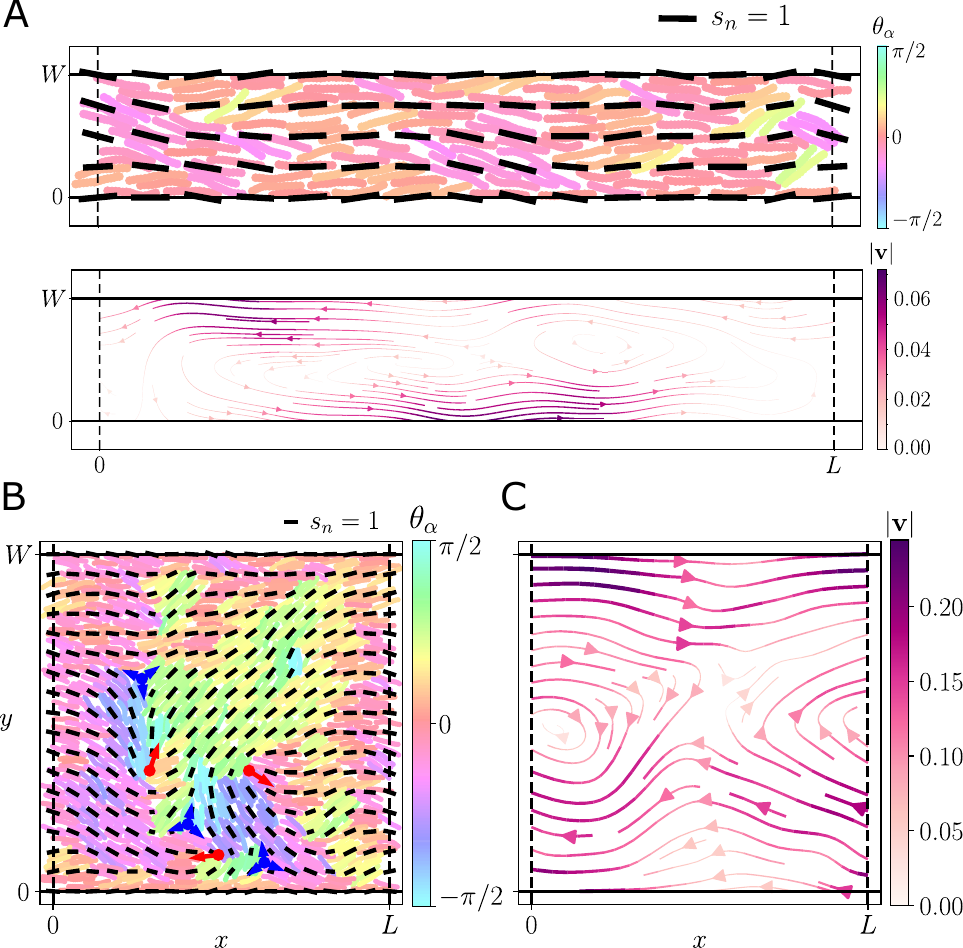}
\caption{\textit{Narrow and large channel width for $v_a=3$.}
\textbf{A}: Snapshots of particle positions and nematic field (top) and velocity field (bottom) for the narrow channel ($N=150$), at $t=1500$.
\textbf{B,C}: Snapshots of particle positions and nematic field (B), velocity field (C) for the large channel ($N=900$), at $t=6000$.
Parameters are $t_{\rm sim}=20000$, $N_{\rm sim}=20$.
}
\label{fig:narrow-large}
\end{figure}

\begin{figure}
\centering
\includegraphics[width=.95\linewidth]{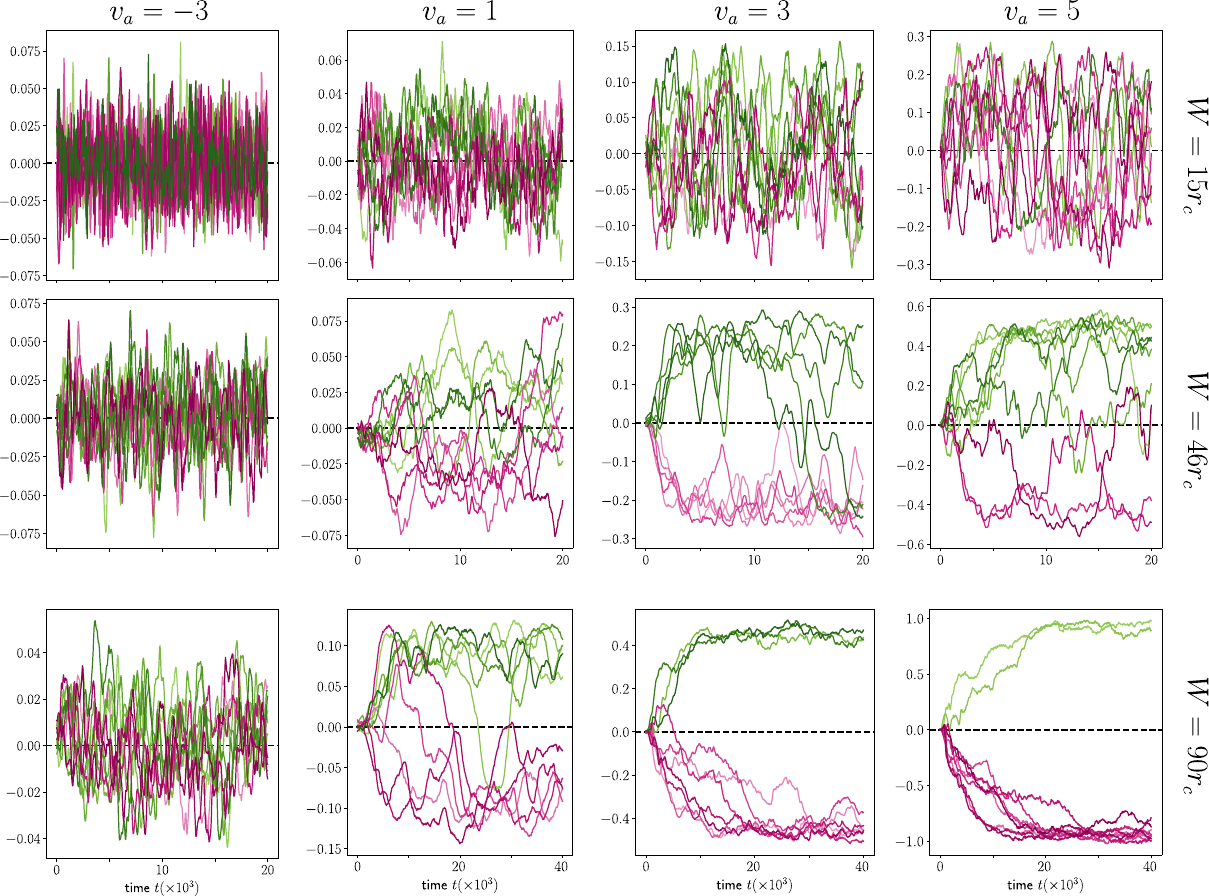}
\caption{\textit{Evolution of linear shear mode $V_x$ in channel geometry.}
Temporal evolution of shear flow $V_x(t)$ for different channel width $W$ (rows) and different activity $v_a$ (columns).
As in Fig.~2 in the main text, green curves satisfy $\langle V_x\rangle_t>0$ and pink curves correspond to $\langle V_x\rangle_t<0$.
Parameters are $N=[150,450,900]$, $t_{\rm sim}=[20-40]\times 10^{3}$, $N_{\rm sim}=20$.
}
\label{fig:vx}
\end{figure}

\begin{figure}
\centering
\includegraphics[width=.95\linewidth]{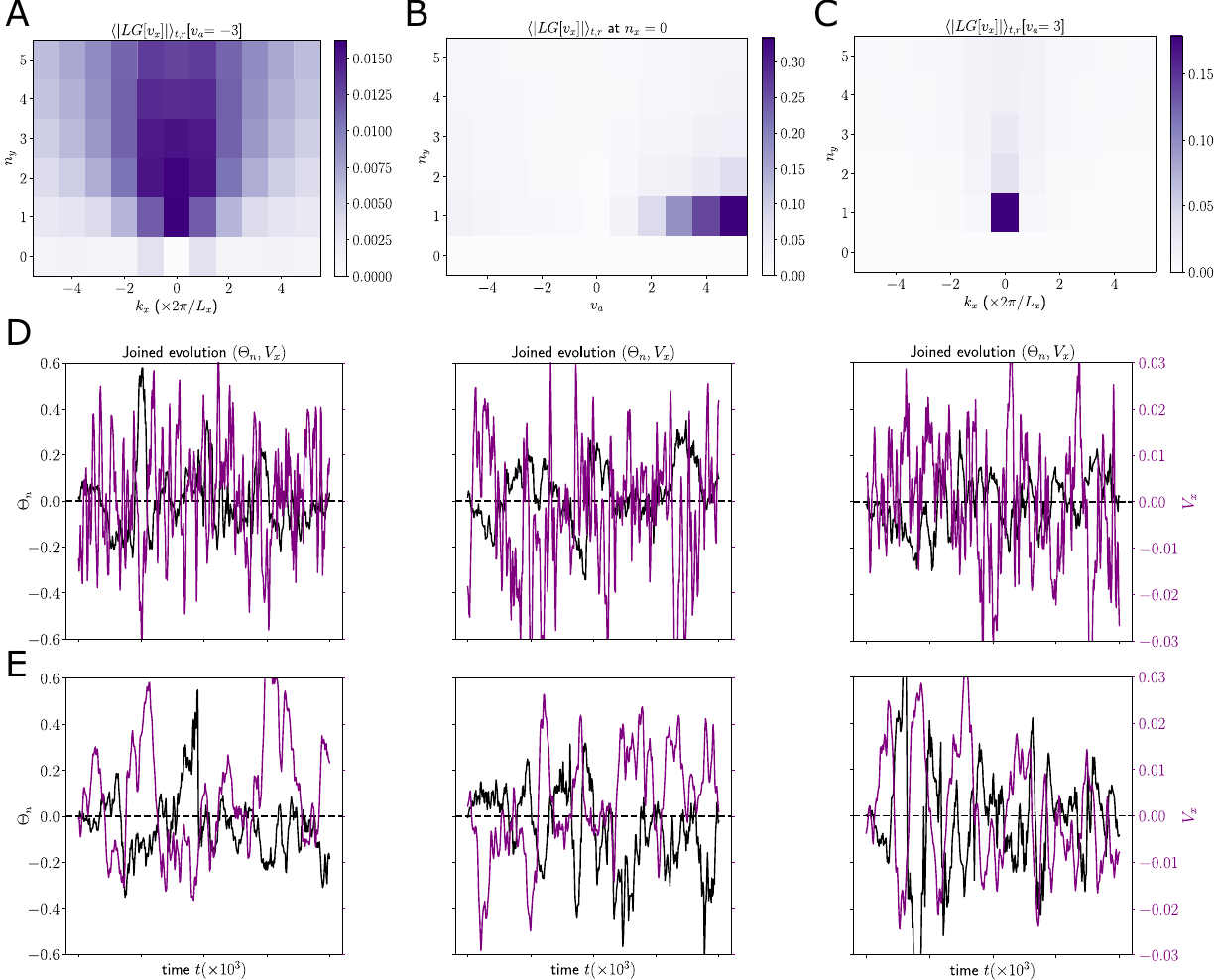}
\caption{\textit{Spectral analysis and flow-orientation anti-correlation in the channel geometry.}
\textbf{A-C}: Average amplitude of Legendre-Fourier projection spectra for $v_a=-3$ (A), $k_x=0$ as a function of activity $v_a$ (B) and $v_a=3$ (C). $n_y$ represents the index of the Legendre polynomial and $k_x=n_x\times 2\pi/L_x$ where $n_x$ is the index of the Fourier basis. See materials and methods for details.
\textbf{D,E}: Joined evolution of shear amplitude $V_x$ and global nematic angle $\Theta_n$ at $v_a=-1$, for narrow channel $W=15r_c$ (D) and intermediate channel width $W=46r_c$ (E). It shows three independent realizations (left to right) with the largest $(V_x,\Theta_n)$ anti-correlation.
Parameters are as in Fig.~3 in the main text.
}
\label{fig:spectrum}
\end{figure}

\begin{figure}
\centering
\includegraphics[width=.8\linewidth]{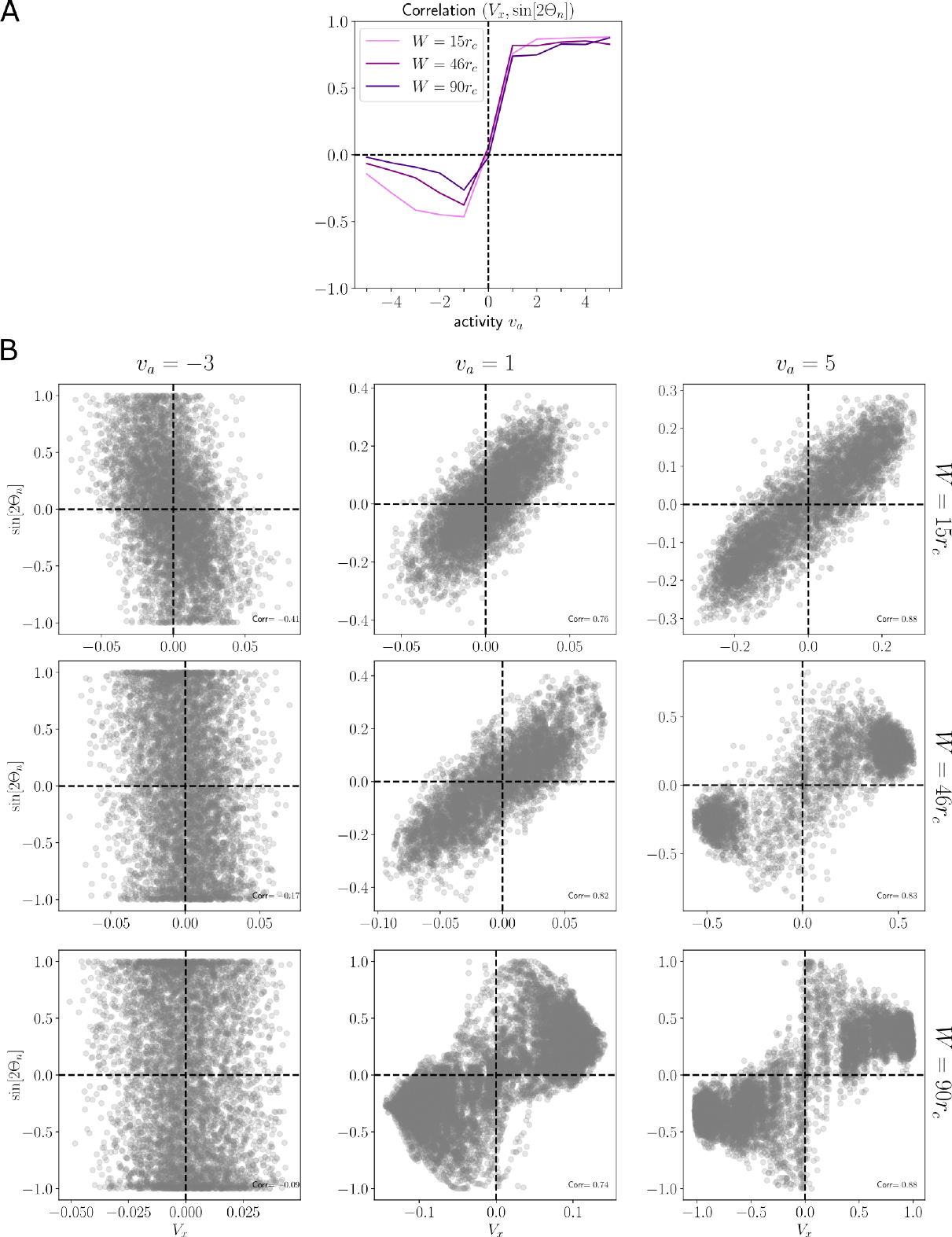}
\caption{\textit{Flow-orientation correlation measures.}
\textbf{A}: Flow-orientation correlation graph as a function of activity $v_a$ and channel width $W$, where Pearson correlation coefficients are computed over all times and realizations as in (B).
\textbf{B}: Scatter plots of $V_x$ and $\sin(2\Theta_n)$ for all times and realizations, varying activity $v_a$ (rows) and channel width $W$ (columns). The Pearson correlation coefficient is indicated on each plot at the bottom right.
Parameters are as in Fig.~3 in the main text.
}
\label{fig:corr}
\end{figure}

\begin{figure}
\centering
\includegraphics[width=.95\linewidth]{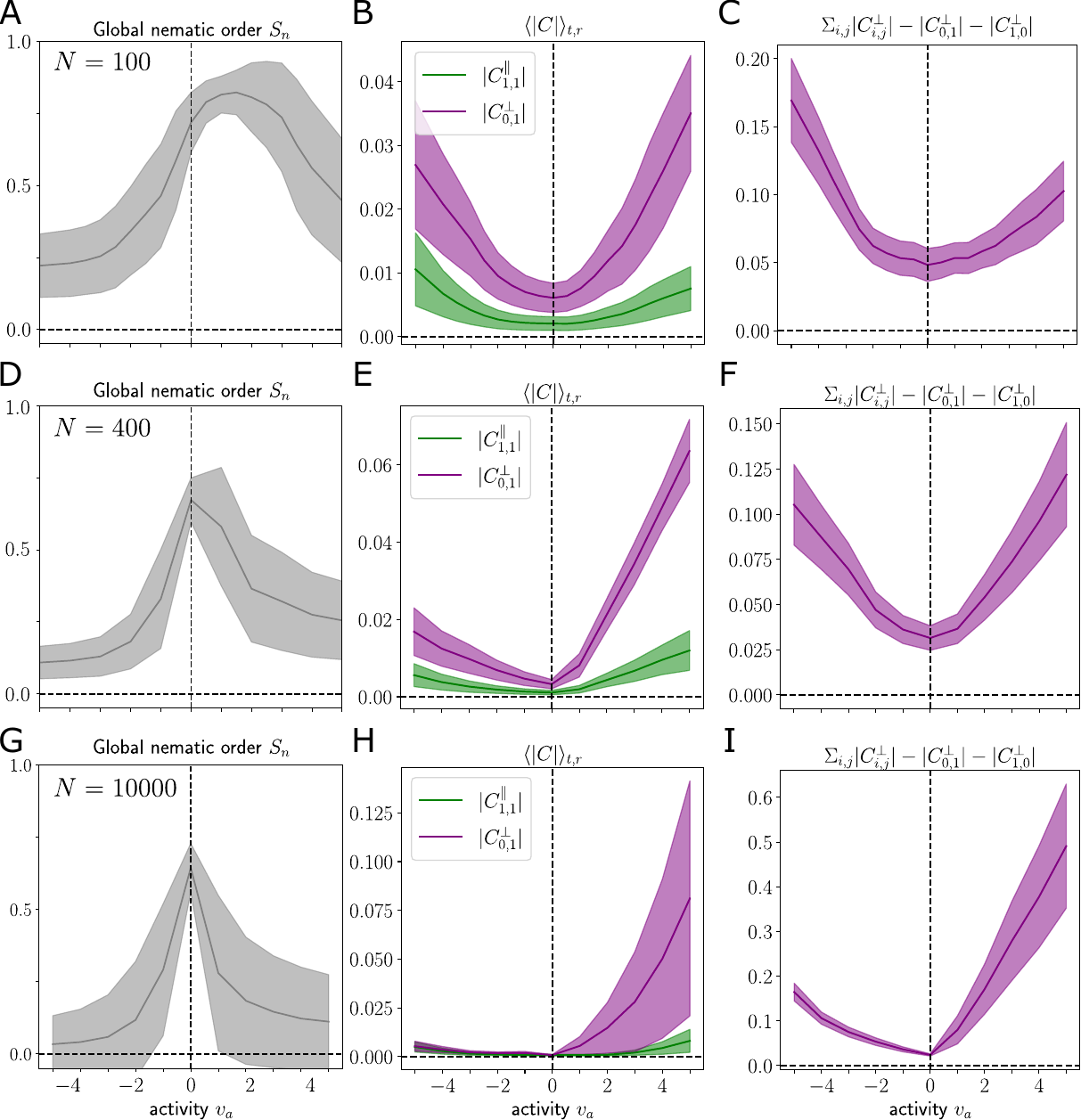}
\caption{\textit{Orientational order and flow transition for PBCs, with $N=100$ (A-C), $N=400$ (D-F) and $N=10000$ (G-I).}
\textbf{A,D,G}: Average global nematic order $S_n$ as a function of activity $v_a$.
\textbf{B,E,H}: Average amplitude of dominant Fourier coefficients as a function of activity $v_a$, with the mode $(0,1)$ for $\mathbf v\cdot\mathbf k_{\perp}$ (pink) and the mode $(1,1)$ for $\mathbf v\cdot\mathbf k$ (green).
\textbf{C,F,I}: Average amplitude of the sum of non-dominant modes as a function of activity $v_a$, for the velocity component $\mathbf v\cdot\mathbf k_{\perp}$.
}
\label{fig:flow-PBC}
\end{figure}

\begin{figure}
\centering
\includegraphics[width=.85\linewidth]{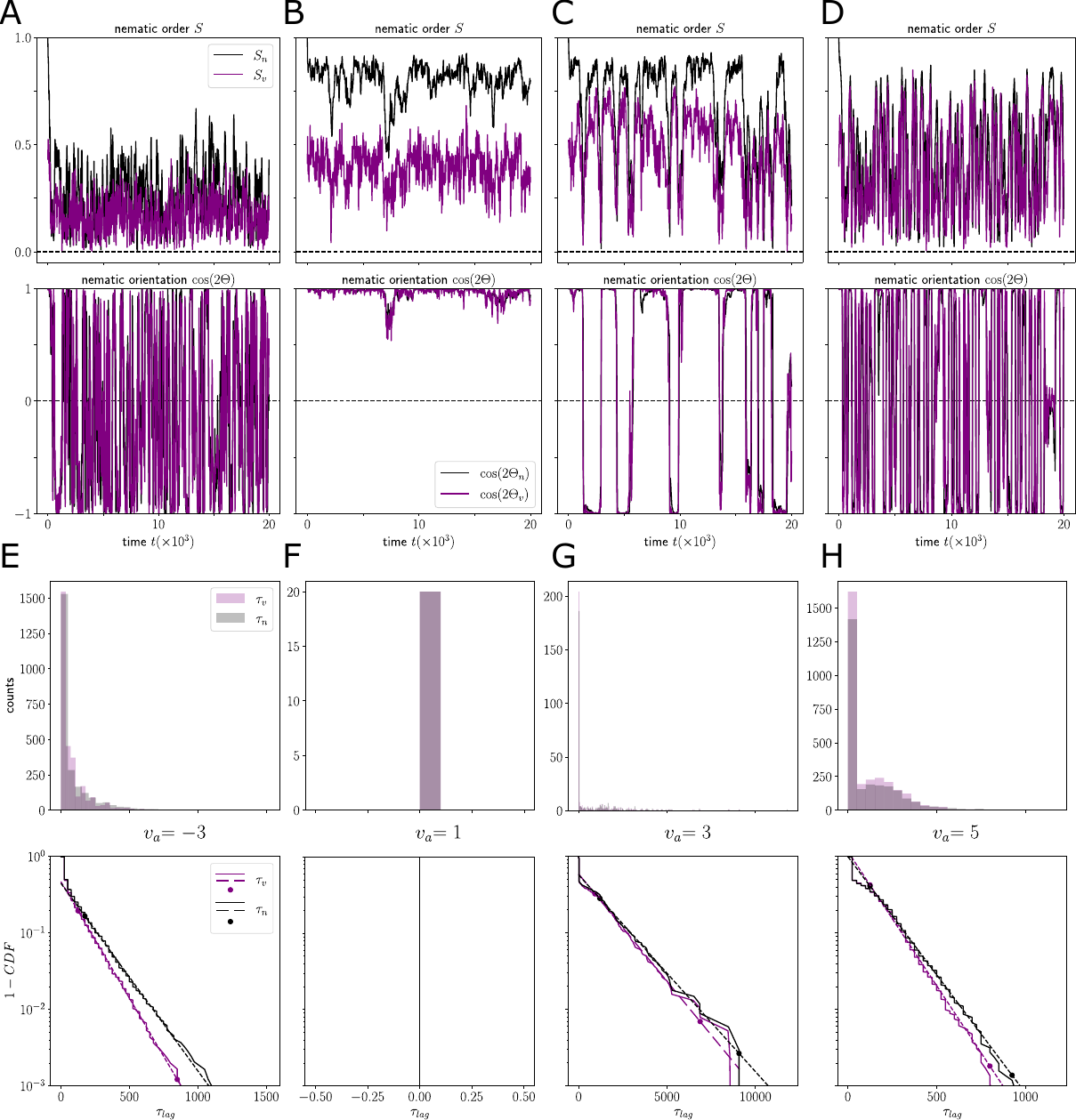}
\caption{\textit{Small active system with periodic boundary conditions.}
\textbf{A-D}: Evolution of the order $S$ (top) and orientation $|\theta|$ (bottom) for individual simulations at $v_a=-3$ (A), $v_a=1$ (B), $v_a=3$ (C) and $v_a=5$ (D).
\textbf{E-H}: Histogram of the distribution of switching times for $|\theta_v|><\pi/4$ (top), and reciprocal cumulative distribution function ($1-CDF$) in log-space (bottom). The dots represent the time interval over which the exponential fit is performed, indicated by a dashed line. Parameters are $N=100$, $t_{\rm sim}=20000$, $N_{\rm sim}=20$.
}
\label{fig:switches}
\end{figure}

Here we consider a two dimensional passive system with channel geometry along the $x$-direction, as described in Figs.~2,3 of the main text for $v_a=0$. We vary the packing fraction $\mathrm{pf}$ and temperature $k_{\rm B}T$, \fig{fig:passive}. Note that for a fixed value of the number of particles per agent $P$, packing fraction $\mathrm{pf}$ and agent density $N/\Omega$ are equivalent. The initial distribution of orientations for agents is horizontal. \fig{fig:passive}A shows a snapshot of the system at low packing fraction, \fig{fig:passive}C at high packing fraction.

First, we compute the global nematic order $S_n$ (defined in materials and methods) on \fig{fig:passive}B. At low temperature, a high packing fraction induces a large nematic order~\cite{Onsager1949}, \fig{fig:passive}C, whereas the nematic order remains small for low $\mathrm{pf}$ \fig{fig:passive}A. At larger temperature, fluctuations are too large to maintain agent-agent alignment and nematic order decreases~\cite{Maier1958}. Thus, the parameters chosen in the main text, $k_{\rm B}T=0.1$ and $\mathrm{pf}=0.8$ correspond to a regime of high nematic order.

To ensure that a high nematic order corresponds to a nematic fluid phase, we compute the mean square displacement (MSD) of agents as a function of packing fraction for temperature $k_{\rm B}T=0.1$, \fig{fig:passive}D. It is defined as
\[
\text{MSD}(\Delta t)=\langle|\delta\mathbf r_p(t_0+\Delta t)-\delta\mathbf r_p(t_0)|^2\rangle_{p,t_0},
\]
averaged over particles $p$ and initial times $t_0$, with $\delta\mathbf r_p=\mathbf r_p-\mathbf r_{\rm com}$ the particle position shifted from the center of mass $\mathbf r_{\rm com}$ of the system.
At long times, the MSD is linear in time, corresponding to diffusive behavior, and becomes larger than the squared agent length over a characteristic time smaller than the simulation time. This indicates that neighbor exchange events occur, as expected in a liquid phase. Note that for temperatures smaller than $k_{\rm B}T=0.1$, the dynamics progressively becomes jammed (data not shown), and we choose parameters to avoid this solid-like phase.

In addition, we compute the auto-correlation functions (ACF) for orientation, $\bar{C}_{uu}$, and velocity, $\bar{C}_{vv}$, using from agents $\alpha$ their orientation $\hat{\mathbf u}_{\alpha}$ and velocity $\mathbf v_{\alpha}$. They are respectively defined as
\[
\bar C_{uu}(\Delta t)=\langle 2[\hat{\mathbf{u}}_{\alpha}(t_0+\Delta t)\cdot\hat{\mathbf{u}}_{\alpha}(t_0)]^2-1\rangle_{\alpha,t_0}
\]
and
\[
\bar C_{vv}(\Delta t)=\langle\mathbf v_{\alpha}(t_0+\Delta t)\cdot\mathbf v_{\alpha}(t_0)\rangle_{\alpha,t_0}/\langle\mathbf v_{\alpha}^2(t)\rangle_{\alpha,t},
\]
averaged over agents $\alpha$ and initial times $t_0$. The orientation ACF relaxes approximately to $S^2$ at equilibrium \cite{Alhissi2024}, and \fig{fig:passive}E confirms that nematic order increases with $\mathrm{pf}$. The relaxation time to a steady-state value occurs over a characteristic time $\tau_u\approx 150$, significantly smaller than the total simulation time. In addition, velocity ACF shows a fast relaxation of velocity correlations~\fig{fig:passive}F, over a characteristic time $\tau_v\approx 20$.

Thus, we confirm that the parameters $\mathrm{pf}=0.8$ and $k_{\rm B}T=0.1$ in the main text correspond to an equilibrium nematic phase without activity. The equilibrium properties of the passive nematic fluid described here can be probed over simulation times much longer than $\tau_u$ and $\tau_v$, hence the choice $t_{\rm sim}\geq 5000$ in the main text or $t_{\rm sim}=2000$ here.

\subsubsection*{Density orientation-coupling for density dipoles at $+1/2$-defect sites}
It is shown on \fig{fig7} of the main text that a positive density dipole $d=(1/A)\int_A\mathrm{d}^2\mathbf r\,[\hat{\mathbf p}\cdot(\mathbf r-\mathbf r_{+1/2})]\hat{\rho}(\mathbf r)/|\mathbf r-\mathbf r_{+1/2}|$ emerges at sites of $+1/2$-defect even in the passive case ($v_a=0$).
Here, we identify a term in the free energy of a nematic liquid crystal that captures this effect.

We start from the free energy of two-dimensional compressible nematics~\cite{Marchetti2013}, expanded around a state of uniform nematic order $\mathbf q(\mathbf r)=\mathbf q_0$ and uniform density $\rho(\mathbf r)=\bar\rho$.
\begin{equation}
\mathcal F=\int_A\mathrm{d}^2\mathbf{r}\,\left[-\frac{\chi_2}{2}\mathbf q:\mathbf q+\frac{\chi_4}{4}(\mathbf q:\mathbf q)^2+\frac{1}{2}K(\hat\rho)|\bm\nabla\mathbf q|^2+w\,\mathbf q:\bm\nabla\bm\nabla\hat\rho+\frac{B}{2}\hat\rho^2\right]
\end{equation}
with reduced density field $\hat{\rho}(\mathbf r)=[\rho(\mathbf r)-\bar\rho]/\bar\rho$. The first two terms favor global nematic order and are irrelevant here. The third term penalizes orientation gradients with a density-dependent elastic constant $K$. The fourth term is a minimal coupling between orientation and density gradients that satisfies nematic symmetry. The fifth term quantifies material compressibility with bulk modulus $B$.

At equilibrium, the configuration that minimizes the free energy $\mathcal F$ is such that the molecular field $\mathbf h=-\delta\mathcal F/\delta\mathbf q$ and the chemical potential $\mu=\delta\mathcal F/\delta\hat\rho$ both vanish. One finds
\begin{align}
\mathbf h&=[\chi_2-\chi_4(\mathbf q:\mathbf q)]\mathbf q+K\Delta\mathbf q-w\bm\nabla\bm\nabla\hat\rho=0 \\
\mu&=\frac{1}{2}\frac{\partial K}{\partial\hat\rho}|\bm\nabla\mathbf q|^2+w\,\bm\nabla\bm\nabla:\mathbf q+B\,\hat\rho=0
\end{align}
Expanding the elastic constant $K$ as $K(\hat\rho)=k_0+k_1\hat\rho+k_2\hat\rho^2+...$, one finds for the equilibrium density
\begin{equation}\label{eq:mu-wk12}
[B+k_2|\bm\nabla\mathbf q|^2]\hat\rho=-w\,\bm\nabla\bm\nabla:\mathbf q-\frac{k_1}{2}|\bm\nabla\mathbf q|^2
\end{equation}
For a given nematic texture, the density dependence of the elastic constant $K$ thus only generates a global shift of the density $\hat\rho\approx-k_1|\bm\nabla\mathbf q|^2/B$ around defects, where $|\bm\nabla\mathbf q|^2$ is large, but not a density dipole.

To understand the effect of the density-orientation coupling with coefficient $w$, we consider the nematic texture of an isolated $+1/2$-defect with director $\hat{\mathbf n}=\cos\theta\mathbf{e}_x+\sin\theta\mathbf{e}_y$. For simplicity, we neglect variation of the nematic order $s\simeq s_0$ and consider the angle $\theta=\arctan_2(y/x)/2$ which minimizes the Frank free energy for a $+1/2$ localized at the origin. Here, the defect orientation is $\hat{\mathbf p}=+\mathbf e_x$ and one computes $q_{xx}=s_0\,x/r$ and $q_{xy}=s_0\,y/r$ where $r=\sqrt{x^2+y^2}$. In this simple case, one obtains $\bm\nabla\bm\nabla:\mathbf q=-s_0\,x/r^3$ such that the equilibrium density in Eq.~\ref{eq:mu-wk12} becomes $\hat\rho=(w\,s_0/B)\,x/r^3$ for $k_1=k_2=0$. Over a ring $r=[R_1,R_2]$ where $s\simeq s_0$, this corresponds to a positive density dipole $d=(w\,s_0/B)\log[R_2/R_1]/(R_2^2-R_1^2)$ if $w>0$, as observed on Fig.~7C in the main text. One interprets $R_1$ as the defect core region where $s_0\sim 0$ whereas $R_2$ defines the dipole region corresponding to $nn$ in figure~\ref{fig:N10000}F.

Alternatively, one could imagine that splay and bend deformations of a nematic material are not symmetric under density variations. This could create a density dipole around $+1/2$-defects having splay (bend) deformations near the tail (head). We thus quantified the elastic energies of splay $E_s=s|\bm\nabla\cdot\hat{\mathbf n}|$ and bend $E_b=s|\bm\nabla\times\hat{\mathbf n}|$ for the director field $\hat{\mathbf n}(\mathbf r)$, weighted by the nematic order field $s(\mathbf r)$. We find a weak anti-correlation with the coarse-grained density field $\hat\rho(\mathbf r)$ for a passive system ($v_a=0$) (figure~\ref{fig:N10000}G). Yet, no bend-splay asymmetry is observed in the passive case, which rules out this mechanism to explain density dipoles at $+1/2$-defects.

In contrast, the density-orientation coupling $f_w=w\,(\mathbf q:\bm\nabla\bm\nabla\hat\rho)$ biases the Hessian of the density $\mathbf H=\bm\nabla\bm\nabla\hat\rho$ with respect to the nematic tensor. Writing $\mathbf q=s(\hat{\mathbf n}\hat{\mathbf n}-\hat{\mathbf n}_{\perp}\hat{\mathbf n}_{\perp})$ with $\hat{\mathbf n}_{\perp}$ a direction perpendicular to the director in the plane, one finds $f=w\,s(H_{\parallel}-H_{\perp})$ with $H_{\parallel}=\hat{\mathbf n}\cdot\mathbf H\cdot\hat{\mathbf n}$ and $H_{\perp}=\hat{\mathbf n}_{\perp}\cdot\mathbf H\cdot\hat{\mathbf n}_{\perp}$. Indeed, our simulations confirm a net bias of the Hessian in those directions, whereas Cartesian components $H_{xx},H_{xy},H_{yy}$ vanish on average (figure~\ref{fig:N10000}H). Finally, computing the two terms of $f_w$, we find in the passive case that $s_n\,H_{\parallel}<0$ whereas $s_n\,H_{\perp}>0$ (figure~\ref{fig:N10000}I) such that $f_w/w=s_n(H_{\parallel}-H_{\perp})<0$ in the passive case $v_a=0$. It demonstrates that density fluctuations are controlled by the nematic textures so as to minimize $f_w$ when $w>0$.

Thus, we can conclude that our simulations confirm the relevance of an orientation-density coupling $f_w=w\,(\mathbf q:\bm\nabla\bm\nabla\hat\rho)$ with $w>0$ for passive compressible systems.


\clearpage
\begin{table} 
\centering
\caption{\textbf{List of parameter values used in the simulations}. Parameters with symbols in parenthesis are only present for a proliferating material. Varied parameters have their values indicated under square brackets. The units are chosen so that $m=r_c=1$ and $\delta t=10^{-3}$.}
\label{table:param}

\begin{tabular}{c c c}
\hline
Parameter & Value & Description \\
\hline\hline
$N_{\rm sim}$ & $[5-20]$ & number of simulations per condition \\
$\delta t$ & $10^{-3}$ & time step \\
$t_{\rm sim}$ & $[1000-40000]$ & simulation time \\
$(k_a)$ & $0$ & rate of agent death \\
\hline
$P$ & $14$ & particles per agent \\
$\mathrm{pf}$ & $0.8$ & $2$D packing fraction \\
$N$ & [100-10000] & number of agents in $2$D \\
\hline
$r_c$ & $1$ & pair potential range \\
$\ell_0$ & $0.5$ & shape link length \\
$(\ell_d)$ & $1.75\,\ell_0(P-1)$ & division length threshold \\
$(u_d)$ & $0.002$ & daughter particle displacement \\
\hline
$m$ & $1$ & particle mass \\
$K_l$ & $[20,50]$ & shape link stiffness \\
$K_w$ & $20$ & external wall stiffness \\
$\kappa_b$ & $[5,17.5]$ & shape bending rigidity \\
$(f_g)$ & $18.75$ & shape division growth force \\
$f_0$ & $2.4$ & inter-agent repulsive coefficient \\
$f_1$ & $0.5$ & inter-agent attractive coefficient \\
$v_a$ & $[-5:5]$ & active flow amplitude \\
\hline
$\xi_s$ & $[5,75]$ & shape dissipative coefficient \\
$\xi$ & $10$ & inter-agent dissipative coefficient \\
$\xi_e$ & $0$ & external (substrate) dissipative coefficient \\
$k_{\rm B}T$ & $[0.05,0.1]$ & temperature (noise) \\
\hline
\end{tabular}
\end{table}


\clearpage

\begin{figure}
\centering
\includegraphics[width=.95\linewidth]{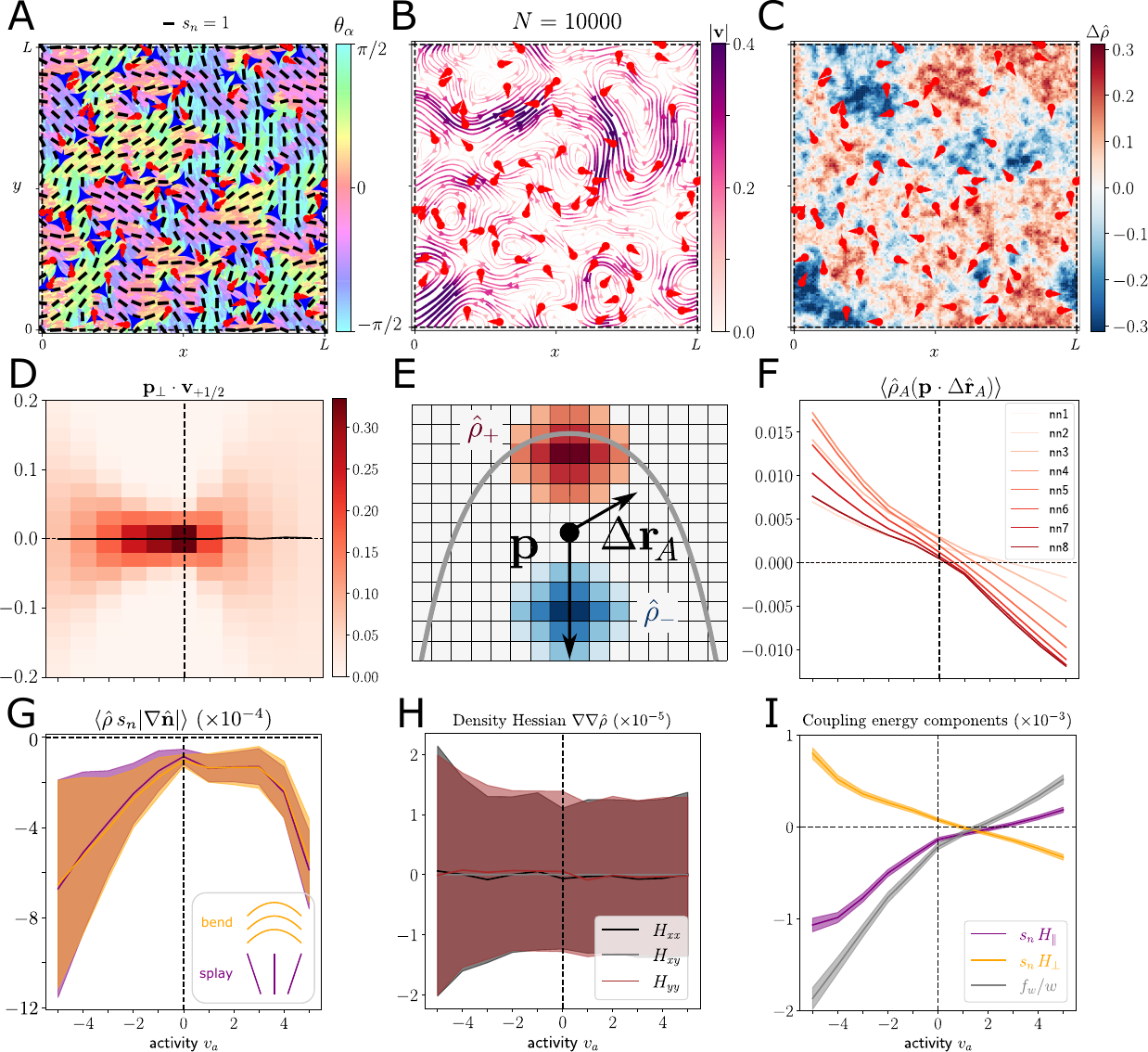}
\end{figure}
\clearpage
\captionof{figure}{\textit{Properties of $\pm 1/2$-defects and density-orientation coupling for a periodic system with $N=10000$.}
\textbf{A-C}: Instantaneous snapshots of the nematic field superposed with particle positions (A), the velocity field (B), and the density variations (C) for extensile activity $v_a=4$. 
\textbf{D}: Histogram as a function of activity of the $+1/2$-defect core velocity $\mathbf v_{+1/2}$ projected in the direction perpendicular to the defect polarity, $\mathbf p_{\perp}=(-\sin\theta_p,\cos\theta_p)$ with $\mathbf p=(\cos\theta_p,\sin\theta_p)$.
\textbf{E}: Sketch for the definition of the density dipole $d=\langle({\mathbf p}\cdot\Delta\hat{\mathbf r}_A)\Delta\hat{\rho}\rangle_A$ with $\Delta\hat{\mathbf r}_A=\Delta\mathbf{r}_A/|\Delta\mathbf{r}_A|$. The black dot indicates the defect center $\mathbf{r}_{+1/2}$, and defines a displacement vector $\Delta\mathbf{r}_A=\mathbf{r}_A-\mathbf{r}_{+1/2}$ around a square patch $A$ centered at $\mathbf{r}_{1/2}$. In this sketch, one has $d<0$. The square patch $A=L^2$ has a length $L=2\mathrm{nn}+1$ in pixel units defined by coarse-grained fields, with $\mathrm{nn}\geq 1$.
\textbf{F}: Average density dipole at sites of $+1/2$-defects as a function of activity, for different next-neighbor patches with area $A=(2\mathrm{nn}+1)^2$ in pixels units (E).
\textbf{G}: Spatio-temporal average of the bend-density correlation $\text{Corr}_b=\langle\hat{\rho}\,s_n|\bm\nabla\times\hat{\mathbf n}|\rangle$ (orange) and the splay-density correlation $\text{Corr}_s=\langle\hat{\rho}\,s_n|\bm\nabla\cdot\hat{\mathbf n}|\rangle$ (purple) as a function of activity $v_a$.
\textbf{H}: Spatio-temporal average of the Cartesian components of the Hessian matrix $\mathbf H=\bm\nabla\bm\nabla\hat\rho$, as a function of activity $v_a$.
\textbf{I}: Spatio-temporal average free energy terms from density-orientation coupling as a function of activity $v_a$, $s_n H_{\parallel}=s_n\partial_{\parallel}^2\hat\rho$ (purple), $s_nH_{\perp}=s_n\partial_{\perp}^2\hat\rho$ (orange), and $f_w/w=s_n(H_\parallel-H_\perp)$ (gray).
}
\label{fig:N10000}

\begin{figure}
\centering
\includegraphics[width=.95\linewidth]{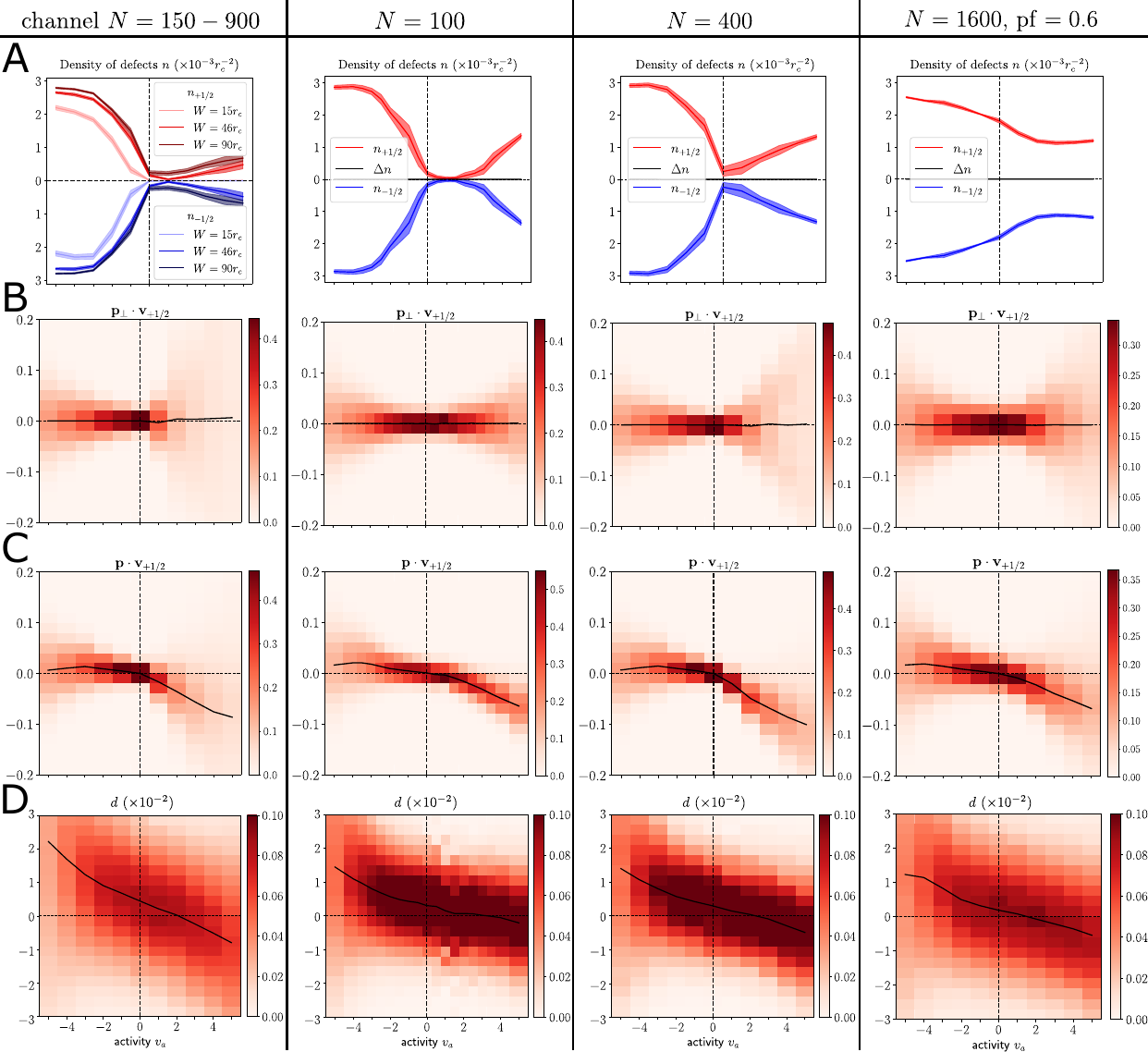}
\end{figure}
\clearpage
\captionof{figure}{\textit{Properties of $\pm 1/2$-defects for different system sizes.}
\textbf{A}: Density of $\pm 1/2$-defects as a function of activity $v_a$.
\textbf{B}: Histogram as a function of activity of the $+1/2$-defect core velocity $\mathbf v_{+1/2}$ projected in the direction perpendicular to the defect polarity, $\mathbf p_{\perp}=(-\sin\theta_p,\cos\theta_p)$ with $\mathbf p=(\cos\theta_p,\sin\theta_p)$.
\textbf{C}: Histogram as a function of activity of the $+1/2$-defect core velocity $\mathbf v_{+1/2}$ projected along the defect polarity $\mathbf p$.
\textbf{D}: Histogram as a function of activity of the density dipole $d$ around the $+1/2$-defect core, with $nn=3$ as in \fig{fig7}C of the main text. 
The left column shows data for channels of widths $W=15r_c$, $46r_c$, and $90r_c$ in (A) and for $W=90r_c$ in (B-D), the middle-left one for a periodic system with $N=100$, the middle-right one for a periodic system with $N=400$, and the right one for a periodic system with $N=1600$.
The packing fraction is $\mathrm{pf}=0.6$ in the right column and $\mathrm{pf}=0.8$ otherwise.
}
\label{fig:defects}


\clearpage 

\subsection*{Movies}
\paragraph{Caption for Movie S1.}
\textbf{Active system in channel geometry for extensile activity.}
Movie showing agent positions (top-left), velocity field (top-right), director field (bottom-left) and relative density field (bottom-right) for parameters $L=92r_c$, $W=46r_c$ and $v_a=3$.
`Animated version of Fig.~\ref{fig1}C.'

\paragraph{Caption for Movie S2.}
\textbf{Active system with small periodic boundaries for extensile activity.}
Movie showing agent positions (top-left), velocity field (top-right), director field (bottom-left) and relative density field (bottom-right) for parameters $L=W=30r_c$ and $v_a=3$.
`Animated version of Figure~\ref{fig4}.'

\paragraph{Caption for Movie S3.}
\textbf{Active system with large periodic boundaries for extensile activity.}
Movie showing agent positions (top-left), velocity field (top-right), director field (bottom-left) and relative density field (bottom-right) for parameters $L=302r_c$ and $v_a=4$.
`Animated version of Figure~\ref{fig6}.'

\paragraph{Caption for Movie S4.}
\textbf{Active system with large periodic boundaries for contractile activity.}
Movie showing agent positions (top-left), velocity field (top-right), director field (bottom-left) and relative density field (bottom-right) for parameters $L=302r_c$ and $v_a=-4$.
`Animated version of Figure~\ref{fig6}.'

\paragraph{Caption for Movie S5.}
\textbf{Three-dimensional active system with side walls for extensile activity.}
Movie showing the coarse-grained velocity field (left) projected in the $xy$-plane at $z=H/2$, and the coarse-grained two-dimensional nematic field superposed with agents satisfying $|z_{\alpha}-H/2|<H/4$. Parameters are $L=W=90.5r_c$, $H=6r_c$, $k_{\rm B}T=0.05$ and $v_a=2$.
`Animated version of Figure~\ref{fig8}A.'

\paragraph{Caption for Movie S6.}
\textbf{Proliferating system with absorbing boundaries for vanishing activity.}
Movie showing agent positions and their orientation (far-left), velocity field (middle-left), director field (middle-right) and relative density field (far-right) for parameters $R_a=49r_c$ (absorbing radius), $k_a=0$, $\ell_d=11.375$, $u_d=0.002$, $f_g=18.75$, $K_l=50$, $\xi_s=75$, $\kappa_b=17.5$ and $v_a=0$.
`Animated version of Figure~\ref{fig8}B.'



\end{document}